\newcommand\pa{\partial}
\newcommand\beq{\begin{equation}}
\newcommand\eeq{\end{equation}}
\newcommand\beqnl{\begin{eqnarray}}
\newcommand\beqna{\begin{eqnarray*}}
\newcommand\eeqna{\end{eqnarray*}}
\newcommand\eeqnl{\end{eqnarray}}

\newcommand\hab{\hat{B}}
\newcommand\has{\hat{S}}

 \def\NN{\hbox{\sf I\kern-.13em\hbox{N}}}
 \def\HH{\hbox{\sf I\kern-.13em\hbox{H}}}
 \def\DD{\hbox{\sf I\kern-.13em\hbox{D}}}
 \def\RR{\hbox{\sf I\kern-.14em\hbox{R}}}
 \def\CC{\hbox{\sf I\kern-.44em\hbox{C}}}
 \def\ZZ{{\hbox{\sf Z\kern-.43emZ}}}
 \def\QQ{\hbox{\sf C\kern -.48emQ}}
 \def\Cc{\hbox{\sf C\kern -.47em {\raise .48ex \hbox{$\scriptscriptstyle |$}}
   \kern-.5em {\raise .48ex \hbox{$\scriptscriptstyle |$}} }}
 \def\Qq{\hbox{\sf Q\kern -.57em {\raise .48ex \hbox{$\scriptscriptstyle |$}}
   \kern-.55em {\raise .48ex \hbox{$\scriptscriptstyle |$}} }}

\documentstyle[prd,aps,multicol,psfig]{revtex}
\begin{document}

\title{ Notes on the third law of thermodynamics.I}
\author{F. Belgiorno\footnote{E-mail address: belgiorno@mi.infn.it}}
\address{Dipartimento di Fisica, Universit\`a degli Studi di Milano, 
Via Celoria 16, 20133 Milano, Italy}
\date{\today}
\maketitle

\begin{abstract}

\vskip -0.3 truecm

We analyze some aspects of the third law of thermodynamics. We first 
review 
both the entropic version (N) and the unattainability version (U) 
and the relation occurring between them. 
Then, we heuristically interpret (N) as a continuity boundary condition 
for thermodynamics at the boundary $T=0$ of the thermodynamic domain. 
On a rigorous mathematical footing, 
we discuss the third law both in Carath\'eodory's approach and in 
Gibbs' one. 
Carath\'eodory's approach is fundamental in 
order to understand the nature of the surface $T=0$. 
In fact, in this approach, under suitable mathematical  
conditions, $T=0$ appears as a leaf of 
the foliation of the thermodynamic manifold associated with 
the non-singular integrable Pfaffian form $\delta Q_{rev}$. 
Being a leaf, it cannot intersect any other leaf $S=$ const. 
of the foliation. We show that (N) is equivalent to the 
requirement that $T=0$ is a leaf. 
In Gibbs' approach, the peculiar nature of $T=0$ 
appears to be less evident because the existence of the entropy 
is a postulate; nevertheless, it is still possible to 
conclude that the lowest value of the entropy 
has to belong to the boundary of the convex set where 
the function is defined.

\vskip -0.3truecm
\end{abstract}

\section{introduction}

We re-analyze the status of the third law of thermodynamics 
in the framework of a purely thermodynamic formalism. 
After a discussion of the status of the third law in current 
physical literature, and after an heuristic justification of 
the entropic version, we set up a rigorous mathematical apparatus in 
order to explore the actual necessity for a third law of thermodynamics. 
The approach by means of Pfaffian forms to thermodynamics, introduced 
by Carath\'eodory, is the most powerful tool for understanding the 
problems which can occur in thermodynamic formalism at $T=0$. 
In our analysis of the latter topic the  
Pfaffian form $\delta Q_{rev}$ is expressed in terms of 
independent extensive variables. One finds that $T=0$, as an integral 
manifold of $\delta Q_{rev}$, can be a leaf of 
the thermodynamic foliation if sufficient regularity conditions 
for the Pfaffian form are ensured. Contrarily, $T=0$ is intersected 
by the (would-be) leaves $S=$ const. which occur at $T>0$. 
The third law appear then as a condition which has to be imposed if 
a foliation of the whole thermodynamic manifold, including the 
adiabatic boundary $T=0$, has to be obtained.\\   
Also Gibbs' approach is analyzed. Carath\'eodory's and Gibbs' approaches  
together allow to better define the problem of the third law.\\ 

\noindent The plan of the paper is the following. 
In 
sect. \ref{law} and in 
sect. \ref{unatta} a discussion of the third law and of its 
standard proofs is given. A particular attention is devoted to 
Landsberg's studies, which are under many respects corner-stones 
on this topic. 
In sect. \ref{plancke} we remark that Planck's restatement of the third 
law is not conventional but mandatory for homogeneous systems.  
In sect. \ref{nhtb}, 
we try to understand, from the physical point of view, 
if it is possible to give a purely thermodynamic 
justification for the third principle in the entropic version (N). 
We show that the third principle in the entropic version 
can be in a natural way interpreted as  a continuity boundary 
condition, in the sense that it corresponds to the natural 
extension of thermodynamics to the states at $T=0$. 
In sect. 
\ref{consist} it is shown that, in the framework of Carath\'eodory 
approach, (N) is equivalent to ensuring that the surface $T=0$ 
is a leaf of the thermodynamic foliation associated with 
the Pfaffian form $\delta Q_{rev}$. 
The isoentropic surfaces cannot intersect the 
$T=0$ surface, because no common point between distinct leaves 
of the foliation determined by $\delta Q_{rev}$ is allowed. 
Some problems arising when (N) is 
violated are discussed, and it is recalled that a singular behavior occur   
if the entropic version (N) fails. 
In sect. \ref{gibbs} a Gibbsian approach to the problem is 
sketched. We show that the entropy can reach its minimum 
value (if any) only on the boundary $T=0$ of its domain.

\section{the third law}
\label{law}

The third law of thermodynamics has been formulated in 
two ways. The original formulation of Nernst concerns 
the behavior of the entropy of every system as the 
absolute zero of the temperature is approached. 
Particularly, the entropic side of Nernst's theorem (N) 
states that, for every system, 
if one considers the 
entropy as a function of the temperature $T$ and of other 
macroscopic parameters $x^1,\ldots,x^n$, the entropy 
difference 
$\Delta_T S\equiv S(T,x^1,\ldots,x^n)-S(T,\bar{x}^1,\ldots,\bar{x}^n)$ 
goes to zero as $T\to 0^+$
\beq
\lim_{T\to 0^{+}} \Delta S =0
\label{nnn}
\eeq
for any choice of $(x^1,\ldots,x^n)$ and of  $(\bar{x}^1,\ldots,\bar{x}^n)$.  
This means that the 
limit $\lim_{T\to 0^{+}} S(T,x^1,\ldots,x^n)$ is a constant 
$S_0$ which does not depend on the macroscopic parameters $x^1,\ldots,x^n$. 
Planck's restatement of (N) is 
\beq
\lim_{T\to 0^{+}} S =0
\label{plancks}
\eeq
and it is trivially 
mandatory for homogeneous systems (cf. sect. \ref{plancke}). 
The other formulation concerns the unattainability 
(U) of the absolute zero of the temperature. 
The (U) side can be expressed as the impossibility 
to reach the absolute zero of the temperature by means of a 
finite number of thermodynamic processes. Both the above formulations 
are due to Nernst, and they are equivalent 
under suitable hypotheses, as it has been remarked 
in Refs. \cite{landsberg,landrmp,landstat} and e.g. also in  
Refs. \cite{haase,wheeler}. 

The third law has a 
non definitively posed status in standard thermodynamics 
and a statistical mechanical basis for it is still 
missing. Counter-Examples to (\ref{plancks}) have been constructed 
\cite{griffiths,lieb}, whereas in Ref. \cite{wald} models displaying a 
violation of (\ref{nnn}) are given. 
Moreover, the validity of thermodynamics for finite-size systems  
if $T$ is sufficiently near the absolute zero has been 
questioned. A corner--stone of this topic is represented by 
Planck's objection (see  
Ref. \cite{simon} and references therein) against a thermodynamic 
description of a ``standard" system below a given temperature, 
due to a reduction of the effective degrees of freedom 
making impossible even to define an entropy. 
The same problem is analyzed in 
Ref. \cite{munster} where the breakdown of thermodynamics 
near the absolute zero is shown in the case of a Debye 
crystal. Thermodynamic formalism is shown to fail because of 
finite size effects. Indeed, if 
the finite size of a real thermodynamic system 
is taken into 
account, according to Ref. \cite{munster} near the absolute zero 
it is no more possible to neglect statistical fluctuations 
in the calculation of thermodynamic quantities like e.g. $T,S$ because 
they are of the same order as the ``standard" leading 
terms\footnote{The example of Ref. \cite{munster} involves a Debye crystal 
having a volume $V\sim 
1$cm$^{3},\langle N \rangle \sim 10^{21}$; statistical fluctuations 
are of the same order as the leading terms for $T\sim 10^{-5}$K.}. 
There is a relative uncertainty in the definition of 
equilibrium states which is of order one. 
Of course, if one considers for the number of degrees 
of freedom a mathematical limit to infinity, then the formal success of the 
thermodynamic approach follows. For more details see Ref. \cite{munster}. 
See also Ref. \cite{wheeler2}. We don't discuss this topic further on 
in this paper.\\ 
In Refs. \cite{munster,munster2} it is proposed, in agreement 
also with the general 
axiomatic approach of Refs. \cite{landsberg,landrmp}, 
that the third law should 
be assumed as the position of a boundary condition for the 
thermodynamic differential equations, whose experimental validation 
is stated in regions above the absolute 
zero. Moreover, according to statements in 
Refs. \cite{landsberg,landstat}, 
the thermodynamic variables on the ``boundary set" of the states 
at absolute zero temperature could be conventionally defined 
as suitable limits (not depending on the path used to approach 
a particular state at $T=0$) of the thermodynamic variables in ``inner 
points" of the thermodynamic configuration space 
and this is proposed as the only 
satisfactory approach to the definition of thermodynamic 
variables at the absolute zero \cite{landsberg,landrmp}.  
To some extent, the application of the thermodynamic equations 
to the absolute zero should be considered as a rather formal 
extrapolation of the theory in a region beyond its confirmed 
domain of validity, and this could be considered as the main 
reason for introducing a new postulate beyond the zeroth, the 
first and the second law \cite{munster2}. In sect. 
\ref{nhtb} we come back on this topic and give an interpretation 
of Nernst Heat Theorem as a ``continuity'' boundary condition 
for thermodynamics at $T=0$.\\
Concluding this section, it is also remarkable that the third principle, 
if considered as an impotence principle in analogy with the first and 
the second principle \cite{buchdahl}, 
in the (U) version simply does not allow 
to get $T=0$, whereas in the (N) version implies also that the work 
produced by an arbitrarily efficient Carnot machine between $T_2>T_1$ 
(that is, a thermal machine with efficiency arbitrarily 
near $1^-$) vanishes as $T_1\to 0^+$ (see Ref. \cite{kestin2}). 
For an extensive discussion upon the third law see also 
Refs. \cite{wilks,beattie,guggenheim}.

\section{Unattainability vs. entropy behavior at $T=0$: Landsberg's 
analysis and standard proof}
\label{unatta}

We start by discussing (U) and (N) in standard 
thermodynamics. The double implication (U)$\Leftrightarrow$(N), 
according to the analysis developed in Refs. \cite{landsberg,landstat}, 
relies on some hypotheses that it is interesting to recall. 

\subsection{(U)$\Rightarrow$(N) in Landsberg's analysis}

A detailed analysis shows that in standard thermodynamics 
unattainability (U) implies (N) if the following conditions are 
satisfied \cite{landsberg,landstat}:\\ 
\\
a) The stability 
condition 
$(\partial S/\partial T)_{x^1,\ldots,x^n}>0$ is satisfied for any 
transformation such that the external parameters (or deformation coordinates) 
$x^1,\ldots,x^n$ are kept fixed; these transformations   
be called isometric 
transformations \cite{buchdahl}.\\ 
\\
This hypothesis is in general ensured by the suitable 
convexity/concavity properties of the thermodynamic potentials and 
is given for ensured in Landsberg's works \cite{landsberg,landrmp,landstat}. 
It is useful to explicit this hypothesis.\\
\\
b) There are no multiple branches in thermodynamic configuration 
space.\\
\\
For the condition b) an equivalent statement is 
``in thermodynamic space no boundary points 
different from the $T=0$ ones occur" \cite{landsberg}, that is, 
no first-order phase transitions are allowed. 
In our setting, this requirement amounts to choosing a 
continuous entropy function. \\
\\
c) There is no discontinuity in thermodynamic properties of the 
system near the absolute zero.\\
\\
In Ref. \cite{landsberg} a careful discussion of the conditions to 
be satisfied in order to ensure (U) is contained. In particular, 
by following Ref. \cite{landsberg}, if a),b),c) hold and moreover (N) 
fails, then $T=0$ is attainable. 
If a),b) and c) hold, then (U) implies 
(N). If a),b) hold and (N) fails, then (U) implies that a discontinuity 
near the absolute zero has to occur, and such a discontinuity has 
to prevent the attainability of $T=0$ (violation of c)) \cite{landsberg}. 
Landsberg makes 
the example of an abrupt divergence in the elastic constants of a 
solid as a conceivable ideal process preventing a solid 
violating (N) to reach 
a zero temperature state by means of quasi--static 
adiabatic volume variations (the hypothesis 
of Ref. \cite{landsberg} is compatible with the vanishing near 
$T=0$ of the (adiabatic) compressibilities that are related 
with elastic constants 
in ordinary thermodynamics; particularly, for standard systems one can 
define the compressibility modulus as the inverse of the compressibility; 
it is proportional to the Young modulus in the case of a solid). 
Anyway, in standard thermodynamics 
a violation of c) is ruled out and is not 
discussed further on in Ref. \cite{landsberg}. Moreover, in standard 
treatment 
of the third law (U) is associated with the impossibility to 
get states at $T>0$ isoentropic to states at $T=0$, so that c) 
is not taken into account. A further discussion is found in the 
following subsections.\\

\begin{figure}[h]
\setlength{\unitlength}{1.0mm}
\begin{picture}(40,40)
\centerline{\psfig{figure=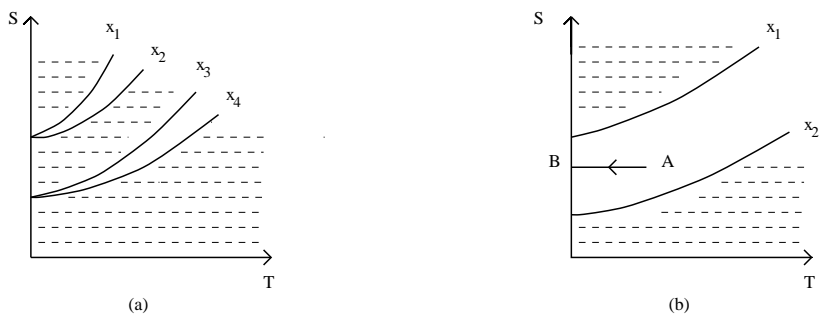,height=4cm,angle=-90}}
\end{picture}
\vspace{0.2cm}
\caption{(a): Multi--branches structure of the thermodynamic space. 
According to Landsberg, 
it implies the validity of (U) and the 
violation of (N). (b): Violation of (N) that implies a violation 
of (U), due to the presence of the isoentropic AB. Landsberg conjectures 
that (U)  
holds if a discontinuity near $T=0$ occurs. See also the text.  
In (a) and (b) the dashed regions are forbidden. }
\label{fig1}
\end{figure}

In Refs. \cite{landsberg,landstat} a further condition  
``entropies don't diverge as $T\to 0^+$" takes into account the 
standard behavior of thermodynamic systems near the absolute zero. 
This condition is not necessary if one considers a non-negative 
concave entropy (cf. sect. \ref{gibbs}). It can be relaxed when 
infinite values of the parameters are allowed \cite{belg31}.
E.g., in the non-standard case of black hole thermodynamics, the 
above condition is not necessary, and in Ref. 
\cite{belgbh} the divergence of the entropy occurring in 
the infinite mass limit  for the black hole case is 
discussed.\\ 
Possible failures of the implication 
(U)$\Rightarrow$(N) are discussed also in Ref. \cite{haase}, 
both in the case of reversible processes and 
in the case of irreversible ones.

\subsection{(N)$\Rightarrow$(U) in Landsberg's analysis}

The isoentropic character of the zero temperature states is 
considered a condition ensuring the unattainability (see e.g. 
Ref. \cite{munster2} and Refs. \cite{landsberg,landrmp}). 
A full implication (N)$\Rightarrow$(U) is possible  
in the case of thermodynamic processes which consist of an 
alternate sequence of quasi-static adiabatic transformations 
and quasi-static isothermal transformations (class P(x) according to 
Refs. \cite{landsberg,landrmp}). 
Actually, a more general notion of unattainability can be 
assumed: ``zero temperature states don't 
occur in the specification of attainable states of 
systems". This is almost literally the (U4) principle as 
in Refs. \cite{landsberg,landrmp}. (U4) states that no process allows 
to reach states at $T=0$, even 
as transient non-equilibrium states. 
Then (N) can fail and (U) can still be valid: In general, the 
latter hypothesis allows a de--linking of (U) and (N) and implies that (N)
$\not \Rightarrow$(U) and (U)$\not \Rightarrow$(N) \cite{landsberg,landrmp}. 
But such a de--linking occurs under particular conditions: the 
failure of the implication (U)$\Rightarrow$(N) requires 
again a rejection of one of the hypotheses b),c) 
above, whereas 
(N)$\Rightarrow$(U) fails if processes not belonging 
to the aforementioned class P(x) allow 
to reach $T=0$ \cite{landsberg,landrmp}. 

For the sake of completeness, we recall in the following subsection 
also the standard approach to Nernst's theorem, which  
involves heat capacities  
\cite{munster2,guggenheim}. 

\subsection{ (U)$\Leftrightarrow$(N) by means of heat capacities: 
the standard proof}
\label{capacities}

The implication (U)$\Rightarrow$(N) can be obtained also 
as follows. It implicitly requires that conditions a),b),c) 
of Landsberg hold. 
Let us consider two  states 
$(T_1,x^1\ldots x^n),(T_2,y^1,\ldots,y^n)$ and the 
related entropies
\beqnl
S(T_1,x^1,\ldots,x^n)&=&S(0,x^1,\ldots,x^n)+
\int_{0}^{T_1}\; \frac{dT}{T}\; C_{x^1,\ldots,x^n}(T)\cr
S(T_2,y^1,\ldots,y^n)&=&S(0,y^1,\ldots,y^n)+\int_{0}^{T_2} 
\frac{dT}{T}\; C_{y^1,\ldots,y^n}(T)
\label{entrzero}
\eeqnl
where $S(0,x^1,\ldots,x^n),S(0,y^1,\ldots,y^n)$ are the 
limits of the above entropies 
as $T\to 0^+$; in the standard proof one assumes also that 
$(T_1,x^1\ldots x^n),(T_2,y^1,\ldots,y^n)$ lie on the same 
isoentropic surface and that it is possible to perform 
a quasi-static adiabatic process connecting them 
\cite{munster2,guggenheim}. If $T_2=0$, then one gets
\beq
S(0,y^1,\ldots,y^n)-S(0,x^1,\ldots,x^n) =
\int_{0}^{T_1}\; \frac{dT}{T}\; C_{x^1,\ldots,x^n}(T) 
\label{uin}
\eeq
and this implies that, if 
$S(0,y^1,\ldots,y^n)-S(0,x^1,\ldots,x^n)>0$ and the stability condition 
$C_{x^1,\ldots,x^n}(T)>0$ holds, a temperature 
$T_1$ satisfying the last equation always exists, so the
unattainability requires 
$S(0,y^1,\ldots,y^n)-S(0,x^1,\ldots,x^n)\leq 0$. The same 
reasoning applied to the process $(T_2,y^1,\ldots,y^n)\to 
(0, x^1\ldots x^n)$ 
gives the opposite inequality 
$S(0,y^1,\ldots,y^n)-S(0,x^1,\ldots,x^n)\geq 0$ and so one 
has to conclude that 
$S(0,y^1,\ldots,y^n)=S(0,x^1,\ldots,x^n)$ \cite{munster2,guggenheim}. 
The convergence of the above integrals of course requires a 
suitable behavior for the heat capacities. 
It is remarkable that, according to this standard proof, (U) is 
implemented by forbidding  
the presence on the same isoentropic surface of states at $T=0$ and 
states at $T>0$. Thus isoentropic transformations reaching $T=0$ cannot exist.
This is a key point. Indeed, one could also allow for a different 
implementation of (U) in which formally states at $T>0$ and states 
at $T=0$ lie on the same isoentropic surface but, because of some 
hindrance arising in a neighborhood of $T=0$, the isoentropic  
transformation reaching $T=0$ actually cannot be performed. 
This is the reason for the hypothesis c) of Landsberg. \\ 

The converse, that is, the implication 
(N)$\Rightarrow$(U), is based on the implicit assumption that 
processes, which don't belong to class P(x) and which allow to 
reach $T=0$, don't exist 
\cite{landsberg,landrmp,landstat}. It is straightforward \cite{guggenheim}: 
Let us consider an adiabatic reversible process 
$(T_1,x^1\ldots x^n)\to (T_2,y^1,\ldots,y^n)$. If (N) holds, 
then there is no possibility to reach a $T=0$ state by means of 
an adiabatic transformation. That is, if (N) holds, then 
(N) and the second law imply (U). 
Indeed, along an adiabatic transformation 
\beq
S(T_1,x^1,\ldots,x^n)=S(T_2,y^1,\ldots,y^n)
\Leftrightarrow 
\int_0^{T_1}\; \frac{dT}{T}\; C_{x^1,\ldots,x^n}(T)= 
\int_0^{T_2}\; \frac{dT}{T}\; C_{y^1,\ldots,y^n}(T)
\eeq
and it is evident that, if the final state is $(T_2=0,y^1,\ldots,y^n)$, then
\beq
\int_0^{T_1}\; \frac{dT}{T}\; C_{x^1,\ldots,x^n}(T)=0,
\label{niu}
\eeq
which is impossible for $C_{x^1,\ldots,x^n}>0$. 
The same conclusion holds if an   
adiabatic transformation from $(T_1,x^1,\ldots,x^n)$ 
to $(0,y^1,\ldots,y^n)$ is considered. 
This proof 
assumes that the only possibility to get $T=0$ is by means 
of a reversible adiabatic transformation. 
The latter is a reasonable 
hypothesis, because any thermal contact and any irreversibility 
cannot be successful in obtaining $T=0$ due to the second law. 
For an interesting proof 
of the above statements see also Ref. \cite{chandler}.

\section{Naive Nernst Heat Theorem: a continuity boundary condition for 
Thermodynamics at $T=0$}
\label{nhtb}

We assume here a physical attitude, and wonder  
if it is possible to give a purely thermodynamic 
justification for the third principle in the entropic version (N). 
This section is dedicated only to an heuristic discussion. 
A  rigorous mathematical setting for the third law 
is found in the following sections.

We stress that, in our reasoning herein, 
we adopt substantially Landsberg's point of view as 
expressed e.g. in Ref. \cite{landstat}, p. 69: ``$\ldots$ 
one must imagine one is approaching the physical situation at $T=0$ 
with an unprejudiced mind, ready to treat a process at $T=0$ 
like any process at $T>0$. With this attitude the maximum information 
concerning conditions at $T=0$ can be deduced $\ldots$''
 
Let us assume that transformations along 
zero temperature states are allowed. In a reversible transformation at $T>0$ 
it is known that $(\delta Q)_{\mathrm rev}= T\; dS.$\\ 
\\
As a consequence,\\
\\
$\Delta S=0$ for adiabatic reversible transformations at $T>0$.\\  
\\
Then, let us consider ideally 
which behavior is natural to postulate 
for thermodynamics at $T=0$.  
Along the $T=0$ isotherm any reversible transformation is 
adiabatic. 
From the point of view of thermodynamic formalism, 
a discontinuity with respect to the natural identification 
between adiabats and isoentropes 
arises if the states at $T=0$ are not assumed to be isoentropic. 
From our point of view, 
the (N) version of Nernst Heat Theorem appears to 
be associated with a ``continuity boundary condition'' 
for thermodynamics at $T=0$. 
Continuity means that the entropy is continuous also on the 
boundary $T=0$ {\sl and} that the identification 
between isoentropic transformations and adiabatic reversible transformations
holds also at $T=0$. 
Indeed, by continuity, it is natural, from 
the point of view of classical thermodynamics, to postulate that 
states at $T=0$ are isoentropic and then in the $T-S$ plane 
the $T=0$ line reduces to a single point 
$(T=0,S={\mathrm const})$. 
But, in order to match continuously 
the property that any isothermal reversible transformation $\gamma_{T=0}$ at 
$T=0$ is isoentropic, 
i.e. $\Delta S=0\; \forall \gamma_{T=0}$, one has to require that 
along the isothermal surfaces the entropy variation becomes smaller 
and smaller, that is, 
$\Delta S_T\equiv S(T,x^1,\ldots,x^n)-S(T,\bar{x}^1,\ldots,\bar{x}^n)$ 
at fixed external parameters has to converge to zero as $T\to 0^{+}$.

\noindent The underlying hypotheses are:\\
\\
$\eta_0$) $T=0$ belongs to the equilibrium thermodynamic phase space;\\
$\eta_1$) it is possible ideally to conceive transformations at $T=0$;\\
$\eta_2$) transformations at $T=0$ are adiabatic reversible;\\ 
$\eta_3$) transformations at $T=0$ are isoentropic;\\
$\eta_4$) there is a continuous match between states at $T=0$ and states 
at $T>0$.\\
\\
Actually, $\eta_4$) could even summarize all the hypotheses above, 
in the sense that a violation of at least one hypothesis 
$\eta_0$),$\eta_1$),$\eta_2$) and $\eta_3$)
would imply a discontinuity in thermodynamics between 
zero temperature states and non-zero ones. 
Concerning $\eta_1$), we recall that Landsberg substantially rejects it, 
because he postulates a poor population of zero temperature 
states in order to forbid the $T=0$ transformation in the Carnot-Nernst cycle. 
Each state can be associated 
with its von Neumann entropy and a priori a violation of (N) and a 
discontinuity are allowed. There is in any case a postulate about the 
density of the zero temperature states which is ``discontinuous'' 
with respect to the assumptions for the states at $T>0$. 

The path of Nernst consists in starting from the violation of the 
Ostwald's formulation of the second law which is implicit 
in a special Carnot cycle, which has the lower isotherm at $T=0$. 
See the figure below. We refer to this cycle as the 
Carnot-Nernst cycle. If it  
were possible to perform it, it would imply the existence of a thermal machine 
with efficiency one, which is a violation of the second law 
of thermodynamics. Note that the violation 
of the identification between adiabats and isoentropes 
is implicit in the $T=0$ isotherm of the Carnot-Nernst cycle. 
In order to avoid this violation, Nernst postulates therefore 
the unattainability (U) of the absolute zero 
(see also Ref. \cite{nernstbook}).

\subsection{transformations at $T=0$}
\label{trasfzero}

Criticisms 
against this path, relating the third law to the second 
one have a long history (see Refs. 
\cite{landrmp,simon,einstein,pippard,kestin2,boas} 
and references therein) which starts with Einstein's objection. 
Einstein underlines that near $T=0$ dissipations begin being  
non-negligible \cite{einstein}. This would make the Carnot-Nernst 
cycle unrealizable because the adiabatic $T=0$ could not be 
performed. 
This kind of criticism could be 
moved also against any attempt to define transformations at $T=0$. 
Nevertheless, it is true that a 
postulate on thermodynamics is required at $T=0$, as variously 
realized in literature (see e.g. Ref. \cite{simon}). 
The objection against the Carnot-Nernst cycle can also avoid referring  
to irreversibility arising near $T=0$, as discussed e.g. in Refs. 
\cite{pippard,boas}. The point is that one reaches 
the $T=0$ surface by means of an adiabatic reversible transformation, 
say BC, and that also any transformation CD 
at $T=0$ has to be adiabatic (see the figure below).

%
\begin{figure}[h]
\setlength{\unitlength}{1.0mm}
\begin{picture}(60,60)
\centerline{\psfig{figure=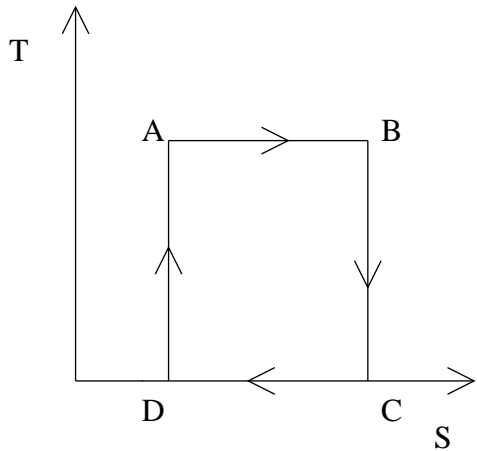,height=6cm,angle=-90}}
\end{picture}
\vspace{0.2cm}
\caption{Carnot-Nernst cycle in the plane $T-S$.}
\label{figcn}
\end{figure}
%

Then, it does not seem possible for the system to be carried along 
the CD transformation contained in the 
$T=0$ surface \cite{pippard,boas} because the adiabatic constraint 
applies to BC as well as to CD and so an operative procedure 
(no matter how ideal) to carry on the cycle seems to be missing. 
It is to be noted that, by analogy with the operative definition 
of isothermal transformation at $T>0$, 
from a physical point of view, the system 
should also be considered to be 
in thermal contact only with a 
``source at $T=0$''. The non-sense, in the 
case of $T=0$, is evident; 
a ``source at $T=0$'' should be defined (a device 
able to exchange large (arbitrary) 
amounts of heat without changing its temperature). 
The point is that any reversible 
transformation at $T=0$ is adiabatic by itself, thus, from 
the point of view of an operative procedure, one should implement 
an adiabatic insulation of the system. 

This kind of reasoning implies  
a failure of the thermodynamic formalism 
at $T=0$, because of the impossibility to give a satisfactory 
prescription for implementing transformations at $T=0$. In 
particular, the problem is related with the existence of 
the intersection between adiabatic surfaces (any isoentropic 
surface intersecting the $T=0$ surface is an adiabatic 
surface which intersects the very peculiar adiabatic surface 
$T=0$), because of the apparent absence of tools allowing 
to pass from one to another one adiabatically. 
In some sense, we find an incompleteness of the thermodynamic 
formalism at $T=0$, because there are serious problems 
in defining an operative procedure \cite{einstein,pippard,boas}. 
Nevertheless, we wish to underline that, in line of principle,  
it could be still possible to implement the adiabatic transformation at $T=0$ 
as a distinct adiabatic transformation , because, even if an adiabatic 
constraint is required, {\sl it corresponds to a path mathematically 
distinguished in the thermodynamic space}. It is clear that, 
if one considers a system described by $(U,V,N)$ and, in 
ideally approaching $T=0$ considers the system as closed, then 
no possibility to distinguish operatively between 
the adiabatic transformation implemented in order 
to approaching $T=0$ and the adiabatic transformation  $T=0$ is left. 
But for systems with a larger thermodynamic space (e.g. systems 
characterized by other deformation parameters) one could have 
closed systems where the adiabatic and isothermal transformation at 
$T=0$ could be implemented. 

A final comment. It is often stated that the third principle 
is not as fundamental as the first and the second ones and 
that it is not related to any new potential in thermodynamics, 
whereas the first law is associated with the internal energy 
function and the second law with the entropy \cite{haase}. 
On this side of the topic, the third law 
prescribes the behavior of the zero-temperature 
part of the entropy $\lim_{T\to 0^+} S(T,x^1,\ldots,x^n )= 
S(0,x^1,\ldots,x^n )\equiv \Sigma (x^1,\ldots,x^n)$ and fixes 
its value to zero. We shall show that\\ 
\\
{\sf The third law of thermodynamics (Planck restatement) 
corresponds to a regularity condition of the Pfaffian equation 
$\delta Q_{\mathrm rev} = 0$ on the boundary $T=0$ of the 
thermodynamic manifold. It is equivalent to the request that 
a well-defined foliation of the whole thermodynamic manifold exists.}\\
\\
In the following sections, we discuss the problem in a 
mathematically rigorous framework.

\section{absolute entropy and Planck's postulate}
\label{plancke}

In our discussion of the third law, the zero-temperature 
entropy constant is undetermined, with the only constraint 
$S_0\geq 0$ suggested by statistical mechanical considerations. 
Planck's restatement of (N) requires $S_0=0$, that is, $S\to 0^+$ as 
$T\to 0^+$, because the constant $S_0$ (entropy at $T=0$), 
which does not depend on the thermodynamic parameters, does not affect 
any physical measurement\cite{planckbook}. 
According to some authoritative experts in the field of thermodynamics, 
this corresponds to a sufficient 
condition for implementing (N), not a necessary one 
\cite{beattie,kestin2,mcglashan}. Problems arising with chemical 
reactions can be suitably solved \cite{beattie,kestin2,mcglashan}. 
An analogous position against the necessity of Planck's restatement 
appears also in statistical mechanics. 
E.g. in Ref. \cite{morandi} statements, according to which systems 
violating $S\to 0^+$ as $T\to 0^+$ in the thermodynamic 
limit, automatically violate (N), have been criticized. 
Residual entropies coming from theoretical calculations in statistical 
mechanics, as far as they are not involved with a 
dependence of the ground state entropy on macroscopic 
parameters, they still cannot be considered as violations 
of the third law \cite{morandi}. 
Also in Ref. \cite{hill} the violation of Planck's statement is 
not considered {\sl a priori} as implying a violation of (N). 
We first discuss the problem {\sl in the framework 
of the thermodynamics of homogeneous systems}; 
then we add 
some comments about the relation with statistical mechanics.\\

Notice that a necessary condition for (N) to hold is that 
$S$ is continuous in the limit $T=0$, whichever state is considered 
on the surface $T=0$. In fact, let us define $X\equiv x^1,\ldots,x^n$ 
and let us assume that $S$ is not continuous 
in $(0,X_0)$. If this discontinuity is not simply an 
eliminable one, there exist two different sequences 
$\{ T_n^{(i)} , X_n^{(i)}\}$, with $i=1,2$, such that 
$(T_n^{(i)} , X_n^{(i)})\to (0,X_0)$ as 
$n\to \infty$ and, moreover, such that
\beq
\lim_{n\to \infty} S(T_n^{(1)} , X_n^{(1)})\not = 
\lim_{n\to \infty} S(T_n^{(2)} , X_n^{(2)}).
\eeq
(N) is badly violated. See also sect. VII. 
The violation of (N) occurs also in the 
case of a (unnatural) eliminable discontinuity. 
As a consequence, the continuity of $S$ on the surface $T=0$ is 
assumed. 

It is to be noted that, if (N) holds, the entropy constant at $T=0$ cannot 
depend on the composition variables $n^i$ which specify the number of 
moles of the component substances which are present in the material 
whose thermodynamic properties are studied. Herein, we let 
composition variables $n^i$ to be included in the set of what 
we called deformation parameters  
\footnote{This choice can be opinable in light of a 
rigorous axiomatic approach \cite{liebyng}, but it allows us to 
call deformation parameters all the parameters different from $U$ (from $T$) 
in our discussion, which is limited to some aspects of the third law.}. 
Then, under 
suitable hypotheses it holds
\beq
\lim_{T\to 0^{+}}\; \frac{\pa S}{\pa n^i}=0.
\eeq
This can be deduced also by means of homogeneity properties 
of the entropy; for a pure phase at constant $p,T$ one has 
\beq
S=n\; \frac{\pa S}{\pa n}.
\eeq
Moreover, one has for a $k$-components system at constant $p,T$
\beqnl
S(T,V,\ldots,n^1,\ldots,n^k)&=&\sum_{i=1}^{k}\; n^i\; 
\frac{\pa S}{\pa n^i}\cr
&\equiv& \sum_{i=1}^{k}\; n^i\; \bar{S}^i
\label{compo}.
\eeqnl
(Note that $\bar{S}^i$ is not the entropy of the single 
component; such an identification would originate a 
wrong expression for the entropy, as it is clear for 
the case of mixtures of ideal gases, where an entropy 
of mixing appears).\\   
As a consequence, it appears that the arbitrary constant 
$S_0$ is zero both for pure phases and for 
mixtures and chemical reactions. In the latter case, the 
third law states that
\beq
\lim_{T\to 0^{+}}\; \sum_{i=1}^{k}\; \nu^i\; \frac{\pa S}{\pa n^i}=0
\eeq
where $\nu^i$ are the stoichiometric coefficients. Actually, 
each derivative should vanish at the absolute zero. 
The alternative definition of Ref. \cite{mcglashan,kestin2} seems 
to be not satisfactory from the point of view of (N), because 
the zero-temperature entropy appears to depend on composition 
variables (which would allow for the composite a different 
zero-temperature entropy for different molar fractions of the components, 
against the postulate of Nernst).\\ 
Then it is thermodynamically appropriate to put $S_0=0$. This 
actually not only does not amount to a real loss of generality, because 
measurements leave the constant undetermined
\footnote{In order to understand this point, it is important 
to underline that 
the constant $S_0$ has actually no operative meaning, in the sense that 
thermodynamic measurements (and extrapolations for the limit 
$T\to 0^{+}$) are relative to the integral of $C/T$. So, in line of 
principle, it can be put equal to $0$ without affecting 
thermodynamic measurements.}; it is also 
a necessity (if the third law holds) for homogeneous systems, 
because $S_0$  
is required to satisfy the 
homogeneity property of the entropy of a homogeneous system. 
$S$ is an homogeneous 
function of degree one in the extensive variables, say, $(U,V,N)$. 
Because of the Euler theorem, this 
implies that, by introducing the operator 
\beq
Y\equiv U\; \frac{\pa}{\pa U}+
V\; \frac{\pa}{\pa V}+
N\; \frac{\pa}{\pa N},
\eeq
the entropy satisfies the equation 
\beq
Y\; S=S.
\label{euho}
\eeq
by inverting 
\beq
T=\left( \frac{\pa S}{\pa U}\right)^{-1}\equiv g(U,V,N)
\eeq
with respect to $U$ [which is allowed by the fact that 
$\pa T/\pa U>0$] one finds $U=h(T,V,N)$ which is an 
homogeneous function of degree one in the extensive variables $(V,N)$. 
Then one obtains 
$S(T,V,N)$ which is homogeneous of degree one 
in the extensive variables $(V,N)$ [it is a quasi-homogeneous 
function of degree one and weights $(0,1,1)$]. If (N) is 
satisfied, then 
\beq
\lim_{T\to 0^+} S(T,V,N)=S_0
\eeq
for any choice of $V,N$. Because of the homogeneity, 
for any $\lambda>0$ one has 
\beq
\lim_{T\to 0^+} S(T,\lambda V,\lambda N)=\lambda 
\lim_{T\to 0^+} S(T,V,N)=\lambda\; S_0
\eeq 
which is consistent with the independence of the limit 
from $V,N$ only for $S_0=0$. No additive constant 
can appear as $T\to 0^+$, because of the homogeneity, thus 
Planck's restatement of the 
third law is mandatory if (N) is satisfied. \\ 
\\
Summarizing:\\ 
\\
{\sf Planck's restatement is mandatory if (N) holds, due to the 
homogeneity property of $S$}.\\
\\
A short comment about the third law in statistical mechanics is 
in order. It is commonly stated that 
a violation of (N) occurs if the ground state is degenerate. 
Moreover, 
As far as the limit 
as $T\to 0^{+}$ is concerned, one has to distinguish between 
finite size systems and bulk systems. 
In the latter 
case, Griffiths shows that, in determining the behavior 
of thermodynamic systems near the absolute zero, in measurements, 
what is really important is the contribution of the excitable 
low-energy quantum states: for bulk systems the 
contribution of the ground state, at reachable low temperatures,  
is irrelevant in determining the behavior of the system, which is instead 
dominated by the contributions of the low-lying energy states 
\cite{griffiths} (the degeneracy of the ground state is not a good 
indicator of the behavior of the entropy for bulk systems at low 
temperature because the thermodynamic limit has to be carried out before 
the limit as $T\to 0^+$ and the two limits in general don't commute 
\cite{griffiths}). In statistical mechanics, 
the ground state degeneracy for bulk systems does not play a 
straightforward role in determining the behavior as the absolute 
zero is approached, 
and examples exist where the ground state is not degenerate 
but the limit $S\to 0^+$ is not implemented \cite{griffiths}. 
However, a role for the degeneracy of the ground state can be 
suitably resorted as in Ref. \cite{lieb}. Therein, 
it is remarked that the entropy functional 
at $T=0$ depends on the boundary conditions. Different boundary 
conditions correspond to different ground states for the bulk system,  
and the contribution of the excitations of the low-lying states 
near the absolute zero can be related with a maximally 
degenerate ground state by means of a variational criterion \cite{lieb}. 
\\  
We limit ourselves to refer the reader to Ref. \cite{narnthi} for a 
further approach to the problem of the third law by means of the 
concept of dynamical entropy and to Ref. 
\cite{callensym} for 
another interesting point of view concerning the problem of the third law 
in presence of ground state degeneracy.

\section{Carath\'eodory's approach and (N)}  
\label{consist}

In Carath\'eodory's approach \cite{caratheodory}, the infinitesimal heat 
exchanged reversibly $\delta Q_{rev}$, defined on a open simply connected 
domain ${\cal D}$, 
is a Pfaffian form, i.e. a one-form $\omega$, whose 
integrability has to be ensured in order to define an entropy function. 
See e.g. Refs. \cite{buchdahl,kestin1,bazarov,boyl1,westenholz}. 
This approach appears to be 
very clarifying with respect to the problem represented by the 
special surface $T=0$. 
In the following, we use $\omega\equiv \delta Q_{rev}$.

\subsection{Foliation in thermodynamics}

Carath\'eodory's {\sl principle of adiabatic inaccessibility} is usually stated 
for the case where ${\cal D}$ has no boundary, that is, $\pa {\cal D}=
\emptyset$ \cite{boyl1,westenholz}.  
It can be formulated as follows:\\
\\
(C): {\sl each neighborhood of any state $x_0$ belonging to the domain 
${\cal D}$ contains states which are inaccessible  
from $x_0$ along solutions of $\omega=0$}.\\
\\
This principles ensures that the Pfaffian form $\omega$ is 
completely integrable, i.e. it satisfies $\omega \wedge d\omega=0$, 
in such a way that a foliation of the thermodynamic manifold into 
isoentropic hypersurfaces is allowed. 
\\
If a boundary is present, there are some changes 
in the theory\footnote{The author is indebted to Lawrence Conlon 
for an enlightening e-mail about the problem of Frobenius theorem 
for manifolds with boundary.}. The integrability condition 
\beq
\omega \wedge d\omega=0
\eeq
has to be imposed in the interior of the domain of the differential 
form $\omega$, where $\omega$ is required to be at least $C^1$. These 
properties ensure that Frobenius theorem can be applied and  
one obtains a foliation in the inner part of 
the manifold. For what concerns the boundary, it can be in part 
transverse and in part tangent to the inner foliation. It is tangent 
when it is a leaf of the foliation itself, i.e. if the boundary 
is an integral manifold for $\omega$ \cite{conlon}. If, instead, 
it is not a leaf, one can induce on the boundary a foliation 
from the inner foliation. Then, a foliation of the whole manifold 
is obtained {\sl if sufficient regularity conditions for 
$\omega$ on the boundary are assumed}.
\\
Let us now consider what happens in thermodynamics. 
The integrating factor $T$ vanishes 
at $T=0$, which means what follows. The non-singular 
integrable Pfaffian form $\delta Q_{rev}$ gives rise 
to a foliation of the thermodynamic manifold for $T>0$. Each leaf 
of the foliation is a solution of the equation $\delta Q_{rev}=0$. 
This foliation 
has codimension one (i.e., each leaf is an hypersurface 
in the thermodynamic manifold). 
For $T>0$, the leaves of 
the foliation are the hypersurfaces $S=$ const. 
One has then to determine if  
the surface $T=0$ is a leaf itself. It is indeed an 
integral submanifold of the Pfaffian form $\omega$, in the 
sense that any curve contained in the surface $T=0$. 
For any initial point lying on the submanifold $T=0$, 
there is a curve $\gamma$ lying entirely in the 
submanifold $T=0$. One has to ensure the uniqueness of the 
solutions of the Cauchy problem for the 
ordinary differential equations associated with 
$\omega$. The Lipschitz condition for each of them would be enough 
in order to get an unique solution.  
If $\omega$ is $C^1$ also on the 
boundary $T=0$, then we can show that the uniqueness is ensured and 
$T=0$ is a leaf of the thermodynamic 
foliation (a tangent leaf). 
The special leaf $T=0$ cannot 
intersect any other leaf $S=$ const. defined at $T>0$, 
because no intersection of leaves is allowed. 
In the following subsections, we analyze the above problem in detail.

\subsection{domain ${\cal D}$}

Let ${\cal D}$ be the thermodynamic manifold whose independent 
coordinates are the extensive variables $U,V,X^1,\ldots,X^n$; 
the variables $V,X^1,\ldots,X^n$ will also be called deformation 
parameters. 
Assume that 
dim${\cal D}=n+2$. ${\cal D}$ is assumed to be a open  
convex set, in order to match the concavity property of $S$.  
Homogeneity requires that 
$(\lambda\; U,\lambda\; V,\lambda\; X^1,\ldots,\lambda\; X^n)$ belongs to 
${\cal D}$ for each real positive $\lambda$, thus ${\cal D}$ has to be 
also closed with respect to multiplication by a positive real scalar, 
i.e., ${\cal D}$ has to be a cone. Then, it is natural to require that 
${\cal D}$ is a convex cone \cite{belhom}. One can also relax 
to some extent the latter condition [e.g., a positive lower bound 
on $V,N$ should be introduced on a physical ground, they cannot 
be arbitrarily near the zero value or statistical fluctuations 
would not allow to define a meaningful thermodynamic state.  
Cf. \cite{belhom}].

\subsection{Pfaffian forms and homogeneous systems} 

Let $\omega\equiv \delta Q_{rev}$ be the 
Pfaffian form of interest, which is identified with the 
infinitesimal heat exchanged reversibly. It is assumed to 
be at least of class $C^1$ in the inner part of the 
thermodynamic manifold. 
One can write 
\beq
\delta Q_{rev}=dU+p\; dV-\sum_i\; \xi_i\; dX^i,
\label{pfafftherm}
\eeq
where $(U,V,X^1,\ldots,X^n)$ are extensive variables. 
The integrability of $\delta Q_{rev}$ ensures that 
\beq
\delta Q_{rev}=T\; dS.
\eeq
We assume that  $\delta Q_{rev}$ is an homogeneous Pfaffian form 
of degree one. This means that the vector field 
\beq
Y \equiv U\; \frac{\pa}{\pa U}+\frac{\pa}{\pa V}+
\sum_i\; X^i\; \frac{\pa}{\pa X^i}
\eeq
is a symmetry for $\delta Q_{rev}$ \cite{cerveau,bocharov}, 
in the sense that 
\beq
L_Y\; \delta Q_{rev} = \delta Q_{rev},
\label{symm}
\eeq
where $L_Y$ is the associated Lie derivative. 
It can be shown that, in the 
homogeneous case \cite{belhom}, an integrating factor  for 
(\ref{pfafftherm}) exists and it is given by
\beq
f\equiv i_Y \delta Q_{rev}=\delta Q_{rev} (Y)=U+p\; V-\sum_i\; \xi_i\; X^i.
\eeq
The integrating factor is required to be such that 
$f\not\equiv 0$, which means that 
$Y$ is not a characteristic or trivial symmetry for the 
distribution associated with $\delta Q_{rev}$. Cf. Ref. 
\cite{bocharov}. Moreover, one requires $f\geq 0$, which is easily shown 
to be equivalent to the conventional choice $T\geq 0$. 
We sketch here some results of Ref. \cite{belhom}. One finds 
\beq
\delta Q_{rev}=f\; d\hat{S}
\eeq
and it can be shown 
in general that, for any homogeneous integrable Pfaffian 
form, $\omega/f$ has to be equal to $dH/H$, where $H$ is 
a positive definite homogeneous function of degree one; 
moreover, the homogeneous 
function $H$ is unique apart from a multiplicative undetermined 
constant\cite{belhom}. This function $H$ is actually the 
entropy $S$, as it can be straightforwardly deduced also 
by direct comparison with the definition of 
$S$ as extensive function
\beq 
S=\frac{1}{T}\; U+\frac{p}{T}\; V-\sum_i\; \frac{\xi_i}{T}\; X^i;
\eeq
in fact, one finds that 
$f$ coincides with the product $T S$.
As a consequence, one has 
\beq
d\hat{S}=\frac{\omega}{f}=\frac{dS}{S},
\eeq
which implies
\beq
S=S_0\; \exp(\int_{\Gamma}\; \frac{\omega}{f}),
\eeq
where $\Gamma$ indicates a path between a reference state 
$U_0,V_0,X^1_0,\ldots,X^n_0$ and the state $U,V,X^1,\ldots,X^n$. 
We require that 
the thermodynamic foliation is described everywhere in ${\cal D}$ 
by the leaves $\hat{S}=$ const., which  means that $\hat{S}$ has 
to be defined everywhere on the thermodynamic manifold (except maybe 
on the boundary $f=0$) \cite{belhom}. The only problems can occur 
where $f=0$. Moreover, one also assumes that to each level set $S=$ const. 
correspond a unique leaf (which means that each isoentropic surface is 
path-connected, as it is natural to assume). 

\subsection{zeroes of the integrating factor and the domain}
\label{zerdom}

Let us define the set 
\beq
Z(f)\equiv \{(U,V,X^1,\ldots,X^n) \; |\; f(U,V,X^1,\ldots,X^n)=0\}. 
\eeq
$Z(f)$ is the set of the zeroes of $f$. We define also 
\beq
Z(T)\equiv \{(U,V,X^1,\ldots,X^n)\; |\; T(U,V,X^1,\ldots,X^n)=0\}, 
\eeq
and
\beq
Z(S)\equiv \{(U,V,X^1,\ldots,X^n)\; |\; S(U,V,X^1,\ldots,X^n)=0\}.
\eeq 
The set $Z(f)=Z(T)\cup Z(S)$ corresponds to an integral manifold of $\omega$ 
($\omega$ is non-singular and $\omega=f d\hat{S}$).

\subsubsection{$Z(T)$}

The set $Z(T)$ is expected to be an hypersurface, but, in general, it 
could be a priori a submanifold of dimension $1\leq k\leq n+1$. 
Actually, it is natural to assume that it is a hypersurface, 
i.e., a manifold of codimension one. 
The equation 
\beq
T(U,V,X^1,\ldots,X^n)=0
\eeq
is required to be implemented for any value of $V,X^1,\ldots,X^n$ which is 
compatible with the system at hand. Contrarily, one should admit that 
$T=0$ could be allowed only for a restricted region of parameters 
(e.g., a crystal could not be allowed to assume a value $V=V_0$ for 
the volume at $T=0$) in such a way that a thermal contact with a lower 
temperature system could not lower the system temperature near the 
absolute zero if values of the parameters outside the allowable range 
would be involved. 
We then assume that the 
$T=0$ is a path-connected hypersurface which coincides with the 
adiabatic boundary $T=0$ of the thermodynamic manifold. 
See also subsect. \ref{bou} .

\subsubsection{Z(S)}

The set $Z(S)$ has to be contained in the boundary of the 
thermodynamic manifold. This is a consequence of the 
concavity of $S$ and of the requirement $S\geq 0$, as 
it is shown in sect. \ref{gibbs}. 
$S=0$ can be moreover attained only on the boundary surface 
$T=0$, in fact 
$S=0$ at $T>0$ can be rejected on physical grounds. 
In fact, any state $z$ such that $T_z>0$ and $S(z)=0$ should 
have the peculiar property to allow the system only to absorb 
heat along any path $\gamma_z$ starting from $z$ in a neighborhood 
$W_z$ of $z$. 
If $C_{\gamma}(T)$ is the heat capacity along a  
path $\gamma$ which does not contain isothermal 
sub-paths, one has that  
\beq
S(y)=\int_{T_z}^{T_y}\; \frac{dT}{T}\; C_{\gamma_z}(T)
\eeq
should be positive for any state $y$ non isoentropic to 
$z$ in $W_z$, 
which is possible only for heat absorption (in fact, 
$C_{\gamma_z}(T)<0$ would be allowed for states such that 
$T_z<T_y$, which would imply heat absorption, and  
$C_{\gamma_z}(T)>0$ would be allowed for states such that 
$T_z>T_y$). The same is true if one considers an isothermal 
path starting at $z$, in fact the heat exchanged would be 
$T_z \Delta S$ and $\Delta S$ should be positive in a neighborhood 
of $z$, being $S=0$ a global minimum of $S$. 
Thermal contact with a colder body at 
$T<T_z$ should allow an heat flow outgoing from the system 
because of the second law in Clausius formulation. Then, 
no possibility to approximate such a thermal contact by means 
of a reversible transformation exists, and this behavior 
can be refused as pathological.\\ 
There is also another argument one can introduce against the 
possibility that, for a non-negative definite entropy, 
the set $Z(S)-Z(T)$ is non-empty. 
By using standard formulas of thermodynamics, 
one has
\beq
S(T,x^1,\ldots,x^{n+1})=S(0,x^1,\ldots,x^{n+1})+
\int_0^T\; \frac{dz}{z}\; C_{x^1,\ldots,x^{n+1}}(z)>0\quad \forall\; T; 
\eeq
$(x^1,\ldots,x^{n+1})$ are deformation parameters (they could be 
also in part intensive); because of the concavity condition 
$C_{x^1,\ldots,x^{n+1}}(T)>0$ the entropy can vanish only for 
$T=0$. A further discussion is found in subsection \ref{bou}. 

Then  
a non-negative concave entropy implies that the set $Z(S)$ 
of the zeroes of $S$ is contained in the set $Z(T)$ of the zeroes 
of $T$:  
\beq
Z(S)\subseteq Z(T).
\eeq
The two sets coincide if (N) holds, otherwise $Z(S)\subset Z(T)$ 
and it could be that $Z(S)=\emptyset$. Then we get the 
following equality:
\beq
Z(f) = Z(T). 
\eeq
If one considers a concave entropy which can be also negative, 
then it happens that $Z(f)\supseteq Z(T)$ because $Z(S)$ is 
not, in general, a subset of $Z(T)$. A typical example is 
the classical ideal gas. Let us consider the monoatomic ideal 
gas. One has \cite{callen} 
\beq
S(U,V,N)=N\; \left[ \frac{5}{2}+\log\left(\frac{U^{3/2}\; V}{N^{5/2}}\; 
\frac{1}{(3\; \pi)^{3/2}}\right) \right];
\eeq
the corresponding Pfaffian form is 
\beq
\omega= dU + \frac{2}{3}\; \frac{U}{V}\; dV+
\frac{2}{3}\; \frac{U}{N}\; \log\left(\frac{U^{3/2}\; V}{N^{5/2}}\; 
\frac{1}{(3\; \pi)^{3/2}}\right)\; dN
\eeq
and one has  
\beq
T=\frac{2}{3}\; \frac{U}{N},
\eeq
and
\beq
f=\frac{2}{3}\; U\; \left[ \frac{5}{2}+\log\left(\frac{U^{3/2}\; V}{N^{5/2}}\; 
\frac{1}{(3\; \pi)^{3/2}}\right) \right]. 
\eeq
In this case one has 
\beq
Z(T)=\{U=0\}
\eeq
and 
\beq
Z(S)=\{ (U,V,N)\; |\; \frac{U^{3/2}\; V}{N^{5/2}}\; 
\frac{1}{(3\; \pi)^{3/2}}=\exp\left(-\frac{5}{2}\right) \}.
\eeq
Then $Z(f)\supset Z(T)$ and $f$ vanishes before $U=0$ is reached.\\

The equation $f=0$ is 
an implicit equation which defines a submanifold of the 
thermodynamic manifold.  This is trivial if $f$ is at least 
$C^1$ everywhere in ${\cal D}\cup \pa {\cal D}$, 
in fact $f=0$ defines a $C^1$ hypersurface 
contained in the domain. This submanifold could be trivially 
an hyperplane $U=U_0=$ const., or a non-trivial hypersurface 
$U=b(U,V,X^1,\ldots,X^n)$. 
A further discussion is found in subsect. \ref{bou} and in 
subsect. \ref{multibranch}.

\subsection{$T=0$ in thermodynamics}

The surface $T=0$ is usually excluded as unphysical. 
In Refs. \cite{liebyng,boyling,borchers} 
an axiomatic approach to thermodynamics excludes a priori 
that the value $T=0$ belongs to the thermodynamic manifold. 
In literature often the claim appears that 
the temperature, for consistency, has to be {\sl strictly} positive 
(see e.g. Refs. \cite{boyling,borchers}).  
From the 
point of view of the approach involving extensive variables, 
it has to be discussed if $Z(T)$ is empty or not (it is 
surely non-empty in the black hole case, see also Ref. \cite{belmac}). 
In the former case, it should be discussed if there is a
lowest temperature \cite{landsberg}
(maybe different for each system) and what 
this implies for the physics. A different lowest temperature 
for each system is not a viable hypothesis, because one could 
put in thermal contact a system at its own lowest temperature with 
another system at a lower temperature and should see an heat 
flux from the former to the latter, and a decrease of the 
temperature of the former. Thus, a lowest temperature should be 
allowed to be the same for all systems.   
Moreover, from an experimental point 
of view there is no apparent limit to the possibility to 
approach $T=0$. From a theoretical point of view, there is 
actually no physical hindrance to consider $T=0$ as a 
possible value. From a mathematical point of view, the Pfaffian form 
$(\delta Q_{rev})$ vanishes at $T=0$ but it is non-singular. 
A singularity of a Pfaffian form $\omega\equiv \sum_i\; a_i(x)\; dx^i$ 
is defined as the set where $a_i(x)=0\ \forall\; i$,  
i.e., where all the coefficients of the Pfaffian form vanish. But 
in the thermodynamic case, no singularity is allowed, because of 
the coefficient of the internal energy term, which is in any case one. 
Thus, no mathematical hindrance to consider $T=0$ in the thermodynamic 
domain appears. The singularity appears only when $T$ is used 
as independent variable, and it is due to the fact that the change of 
variable $U\mapsto T$ is a diffeomorphism only for $T>0$. See also 
the following subsection. This topic is also discussed in Ref. \cite{belg31}.

\subsection{boundary revisited}
\label{bou}

In thermodynamics, as discussed in sect. \ref{zerdom}, 
it is to some extent natural to 
assume that the boundary $T=0$ is described explicitly by a (maybe 
smooth, let us assume at least $C^1$) function:
\beq
U = b(X^1,\ldots,X^{n+1});
\eeq
one can figure that it 
corresponds to the equation for the ground-state energy 
of the system as a function of the deformation parameters, as it 
is clear from the following analysis. 
$b$ is a function which is homogeneous of degree one 
with respect to $(X^1,\ldots,X^{n+1})$:
\beq
b(\lambda\; X^1,\ldots,\lambda\; X^{n+1})=\lambda\; b(X^1,\ldots,X^{n+1}).
\eeq
Thus, $b$ has to be defined on a cone ${\cal K}_b\subset \RR^{n+1}$.  
Moreover, if $U_0,X^1,\ldots,X^{n+1}$ belongs to the boundary $T=0$, 
from 
\beq
T(U,X^1,\ldots,X^{n+1})=\int_{U_0}^U\; dU\; \frac{\pa T}{\pa U}\; 
(U,X^1,\ldots,X^{n+1}),
\eeq
where the integral is an improper integral, because 
$\pa T/\pa U=1/C_{X^1,\ldots,X^{n+1}}\to \infty$ as $T\to 0^+$, 
and from the concavity of $S$, which implies that $C_{X^1,\ldots,X^{n+1}}>0$, 
one finds that $U\geq U_0$, i.e. it has to hold 
$U\geq b(X^1,\ldots,X^{n+1})$. Thus, the domain ${\cal D}$ has to 
be such that the inequality $U\geq b(X^1,\ldots,X^{n+1})$ is implemented 
for each $U$ and for each $(X^1,\ldots,X^{n+1})\in {\cal K}_b$. 
The domain ${\cal D}$ contains the set 
\beq
{\mathrm epi}(b)\equiv \{ (U,V,X^1,\ldots,X^n)\; |\; (V,X^1,\ldots,X^{n+1})\in 
{\cal K}_b, U\geq b(V,X^1,\ldots,X^n) \}. 
\eeq
This set is the so-called epigraph of the function $b$. 
If the function $b(V,X^1,\ldots,X^n)$ 
is required to be convex, then it is defined 
on the convex cone ${\cal K}_b$, and its epigraph ${\mathrm epi}(b)$ 
is a convex cone 
(the epigraph of an homogeneous $b$ is a cone). 
Then,  the domain ${\cal D}$ can be chosen to be 
\beq
{\cal D}={\mathrm epi}(b).
\eeq  
One can also assume that ${\cal D}$ is a convex cone of the form 
\beq
{\cal D}=\{ (U,V,X^1,\ldots,X^n)\; |\; (V,X^1,\ldots,X^{n+1})\in 
I_V\times I_{X^1}\times \ldots I_{X^n},\ U\geq b(V,X^1,\ldots,X^n) \},
\eeq
where the intervals $I_V,I_{X^1},\ldots,I_{X^n}$ are $\RR_+$. 
We can find a coordinatization of the boundary by 
means of coordinates $(B,X^1,\ldots,X^{n+1})$ such that the boundary $T=0$ 
coincides with $B=0$. In fact, we can simply define 
\beq
B\equiv U- b(X^1,\ldots,X^{n+1});
\label{cobu}
\eeq
$B\geq 0$ is a degree one homogeneous function, and $\pa U/\pa B = 1$. 
By inverting one finds 
\beq
U=B+ b(X^1,\ldots,X^{n+1}).
\label{coub}
\eeq
As a consequence, one gets
\beq
\bar{f}\equiv f(B,X^1,\ldots,X^{n+1})=B + b(X^1,\ldots,X^{n+1})-\sum_k\; \xi_k\; X^k;
\eeq
by definition, $\bar{f}$ vanishes for $B=0$, i.e.
\beqnl
&&0=b(X^1,\ldots,X^{n+1})-\sum_k\; \xi_k (0,X^1,\ldots,X^{n+1})\; X^k\\
&&\Leftrightarrow\cr
&&\frac{\pa b}{\pa X^k}=\xi_k (0,X^1,\ldots,X^{n+1})\quad \forall k.
\eeqnl
Notice that, by defining for all $i=1,\ldots,n+1$
\beq
\tilde{\xi}_i (B,X^1,\ldots,X^{n+1}) \equiv \xi_i (B,X^1,\ldots,X^{n+1})-
\frac{\pa U}{\pa X^i}(B,X^1,\ldots,X^{n+1}), 
\eeq
one finds 
\beq
\omega=dB-\sum_i\; \tilde{\xi}_i (B,X^1,\ldots,X^{n+1})\; dX^i, 
\eeq
and it holds 
$\tilde{\xi}_i (B=0,X^1,\ldots,X^{n+1})=0$ for all $i=1,\ldots,n+1$, 
because of the definition for $f=0$ to be an integral hypersurface for 
$\omega$. Moreover, notice that, 
under this assumption about the boundary $T=0$, one 
obtain that $Z(S)\subseteq Z(T)$ necessarily.  
In fact, one can write for an 
everywhere continuous entropy  
\beq
S(B,X^1,\ldots,X^{n+1})=S(0,X^1,\ldots,X^{n+1})+
\int_0^B\; dY\; \frac{1}{T(Y,X^1,\ldots,X^{n+1})}, 
\label{esseb}
\eeq
where $S(0,X^1,\ldots,X^{n+1})$ is the value attained by $S$ at $B=0$ 
by continuity;  
it is evident that $S$ cannot vanish outside $Z(T)$, because 
$S(0,X^1,\ldots,X^{n+1})\geq 0$ and 
$\int_0^B\; dY\; 1/T(Y,X^1,\ldots,X^{n+1})>0$ for all $B>0$.\\

As far as the entropy $S$ as a function of $B,X^1,\ldots,X^{n+1}$ 
is concerned, it is such that
\beqnl
\frac{\pa S}{\pa B}&=&\frac{1}{T(B,X^1,\ldots,X^{n+1})},\\
\frac{\pa S}{\pa X^i}&=&-
\frac{\tilde{\xi}_i (B,X^1,\ldots,X^{n+1})}{T(B,X^1,\ldots,X^{n+1})} 
\quad \forall i=1,\ldots,n+1.
\eeqnl

One could also allow for different choices of coordinates. Instead of $B$ 
defined as above, one could introduce another (maybe local) coordinate for the 
boundary $f=0$, say $\hat{B}$, such that the boundary coincides with $\hab=0$. 
The coordinate transformation $U\mapsto \hab$ 
is required to be regular, i.e.
\beqnl
&&\frac{\pa U}{\pa \hab}\not =0\quad \hbox{for}\ \hab=0\\
&&\cr
&&\frac{\pa \hab}{\pa U}\not =0\quad \hbox{for}\ \hab=0,
\eeqnl
then one can find 
\beq
\omega = \frac{\pa U}{\pa \hab}\; d\hab
- \sum_i \hat{\xi}_i (\hab,X^1,\ldots,X^{n+1})\; dX^i,
\eeq
where
\beq
\hat{\xi}_i (\hab,X^1,\ldots,X^{n+1}) \equiv \xi_i (\hab,X^1,\ldots,X^{n+1})-
\frac{\pa U}{\pa X^i}(\hab,X^1,\ldots,X^{n+1}), 
\eeq
and 
\beq
\hat{f} = \frac{\pa U}{\pa \hab}\; \hab- \sum_i \hat{\xi}_i 
(\hab,X^1,\ldots,X^{n+1})\; X^i.
\eeq
It is important to point out that, also in these coordinates, 
one has
\beq
\hat{\xi}_i (\hab=0,X^1,\ldots,X^{n+1}) = 0. 
\eeq
In the following, we define
\beq
\frac{\pa U}{\pa \hab}\equiv a(\hab,X^1,\ldots,X^{n+1}).
\eeq
Notice that, because of the properties of the Pfaffian form 
$\omega$, the absolute temperature $T$ cannot be used as a good 
coordinate for  
the boundary, in fact $\pa U/\pa T \to 0$ as $T\to 0^+$ for 
all physical systems allowing a finite $S$ at $T=0$. This 
choice (as well as the choice of $f$) 
seems to transform the regular Pfaffian form $\omega$ into a singular 
one, but this trouble is simply due to the singularity 
in the jacobian of the coordinate transformation $U\mapsto T$, which is 
a diffeomorphism only for $T>0$. See also 
\cite{belg31}.

\subsection{condition to be satisfied in order that $T=0$ is a leaf}
\label{leafzero}

In order to understand better the problem of the boundary $T=0$, 
it is useful to recall the equivalence between the equation 
$\omega=0$ and the so-called Mayer-Lie system of partial differential 
equations [herein, $X^i$ stays for any extensive variable different 
from $U$ and $\xi_i$ for the corresponding intensive variable] 
\beq
\frac{\pa U}{\pa X^i}(X^1,\ldots,X^{n+1})=\xi_i (U,X^1,\ldots,X^{n+1})\quad 
\hbox{for}\ i=1,\ldots,n+1.
\label{mayl}
\eeq
One can also assign an initial condition 
\beq
U(X^1_0,\ldots,X^{n+1}_0)=U_0
\label{caul}
\eeq
and thus define a Cauchy problem for the above Mayer-Lie system.  
The integrability condition $\omega \wedge d\omega=0$ in the inner 
part of the manifold is sufficient for a $C^1$ Pfaffian form in order 
to ensure the existence and the uniqueness of the above Cauchy problem. 
This means that the Cauchy problem with 
initial point on the $T=0$ boundary allows solutions which 
lie in $T=0$. If $\omega$ is $C^1$ also on the boundary, then it 
is evident that the aforementioned curves are the only possible 
solutions to the above Cauchy problem with initial point on the surface 
$T=0$. In other terms, if $\omega\in C^1$ everywhere, then 
$T=0$ is a leaf of the thermodynamic foliation. But, a priori, one can 
consider also a Pfaffian form $\omega$ such that it is continuous 
on the boundary $T=0$ but non-necessarily $C^1$ there. The uniqueness  
of the solution of (\ref{mayl}) with initial condition on the 
surface $T=0$ could be ensured if the functions 
$\xi_i (U,X^1,\ldots,X^{n+1})$ 
are locally Lipschitzian with respect to $U$ uniformly with respect to 
$X^1,\ldots,X^{n+1}$ 
in a neighborhood of $(U_0,X_0^1,\ldots,X_0^{n+1})$. 
If even this condition fails, then 
the continuity of $\omega$, i.e., the continuity of $\xi$ also 
in $T=0$ can allow multiple solutions of the differential 
equation (\ref{mayl}).  

Let us consider the following differential equation which describes 
isoentropic curves in the special coordinated adapted to the boundary 
introduced in the previous subsection:
\beq
\frac{dB}{d\tau}=\sum_i\; \tilde{\xi}_i 
(B(\tau),X^1(\tau),\ldots,X^{n+1} (\tau)) 
\frac{dX^i}{d\tau}
\label{leafeq}
\eeq
This equation can be easily obtained from (\ref{mayl}). 
Let us consider at least piecewise $C^1$ 
functions $X^1(\tau),\ldots,X^{n+1} (\tau)$ 
for $\tau\in [0,1]$. These functions are arbitrarily assigned. 
Let us consider a solution curve such that 
$\lim_{\tau \to \tau_0} B(\tau)=0$ for $\tau_0\in [0,1]$. By hypothesis, 
$X^1(\tau),\ldots,X^{n+1} (\tau)$ are finite for $\tau\to \tau_0$. 
Then, by continuity, such a solution can 
be extended to $\tau=\tau_0$, i.e. $T=0$ cannot be a leaf. 
This happens as a 
consequence of well-known theorems on the ordinary differential equations,  
see \cite{walter}, pp. 67-68. 
Let us then consider a point 
$(B(0)=0,X^1(0)=X^1_0,\ldots,X^{n+1} (0)=X^{n+1}_0)$ on the surface $T=0$. 
In each neighborhood of this point, one can find inner points, each of 
which belongs to a surface $S=$ const. In fact, 
each point of the surface $B=0$ is a limit point for the nearby inner 
points of the thermodynamic domain and each inner point has to belong 
to a $S=$ const. integral manifold, because of the integrability condition. 
If $T=0$ is a leaf [or if the connected components of $T^{-1}(0)$ are 
leaves if $T=0$ is not connected], the only 
possibility is that, in approaching $T=0$, one is forced to change 
leaf $S=$ const., i.e., it is not possible to approach $T=0$ by 
remaining on the same leaf $S=$ const., otherwise the inner solution 
of $\omega=0$ could be extended to $T=0$. 
Notice that, in case there exist two solutions of the Cauchy 
problem for the differential equation (\ref{leafeq}) 
with initial condition $(B(0)=0,X^1(0)=X^1_0,\ldots,X^{n+1} (0)=X^{n+1}_0)$, 
one lying in the $T=0$ surface and the other leaving the $T=0$ 
surface, then these solutions are tangent at the initial point. This 
means that, in case of existence of the limit as $B\to 0^+$ 
of the entropy, when (N) is violated, there are surfaces $S=$ const. 
which are tangent to the submanifold $T=0$. We now prove these 
statements.

\subsubsection{validity of (N)}

If (N) holds, 
then $T=0$ plays at most 
the role of asymptotic manifold for the inner leaves $S=$ const. 
No inner integral manifold can intersect $T=0$, 
i.e., $T=0$ is a leaf. In order to 
approach $T=0$ at finite deformation parameters, one has necessarily to 
change from one adiabatic surface $S=$ const. to another one, 
it is impossible to approach $T=0$ by means of a single adiabatic 
transformation. We have then a mathematical explanation of the 
naive unattainability picture sketched by means of $S-T$ diagrams 
one finds in standard textbooks on thermodynamics (cf. also 
the definition of $P(x)$ transformations in \cite{landsberg}).

\subsubsection{violation of (N)}

If, instead, (N) is violated and 
$\lim_{B\to 0^+}\; S(B,X^1_0,\ldots,X^{n+1}_0)\equiv 
S(0,X^1_0,\ldots,X^{n+1}_0)$, then the inner integral manifold 
$0<S(0,X^1_0,\ldots,X^{n+1}_0)=$ const. exists and can be continuously 
extended to $T=0$.\\ 
This can be proved by means of a variant of the 
implicit function theorem. For simplicity, we put here 
\beq
X\equiv X^1,\ldots,X^{n+1}. 
\eeq
Let us consider a point $(0,X_0)$ 
which is not a local minimum for $S$, i.e. it is not such that 
$S(0,X_0)\leq S(B,X)$ in a neighborhood of $(0,X_0)$. Such a point 
surely exists if (N) is violated, as it is easy to show (cf. also 
\cite{belg31}).  
We are interested in the zeroes of the function 
\beq
\sigma (B,X)\equiv S(B,X)-S(0,X_0). 
\eeq
The function $\sigma (B,X)$ is a continuous function 
which is monotonically strictly 
increasing in $B$ everywhere in the domain ${\cal D}\cup \pa {\cal D}$, 
because $S(B,X)$ is, by construction, a strictly 
increasing monotone function in $B$, as it is evident from 
(\ref{esseb}).\\ 
In particular, we wish to know if there is a continuous function $B(X)$ 
defined in a neighborhood of $(0,X_0)$  
such that $\sigma (B(X),X)=0$ and such that $B(X_0)=0$. If $(0,X_0)$ 
is a strict local minimum for $S$, then by definition there exists a 
neighborhood of $(0,X_0)$ where $\sigma (B,X)>0$, thus the aforementioned 
function $B(X)$ does not exists. If it is a weak local minimum, 
in the sense that $S(0,X_0)\leq S(B,X)$ in a neighborhood and the 
equality is allowed, again $\sigma (B(X),X)=0$ does not admit solutions, 
in fact if $S(B,X)=S(0,X_0)$ is allowed in a neighborhood 
$W\supset [0,B_0)\times V$ of $(0,X_0)$, with $V\ni X_0$ open set, 
then $S(0,X)<S(0,X_0)$ because $S(0,X)< S(B,X)$. Thus, being $(0,X_0)$ 
a local minimum, one can find a smaller neighborhood where 
$S(B,X)=S(0,X_0)$ is impossible for any $B>0$ ($S(0,X)=S(0,X_0)$ is 
instead allowed).\\
Notice that $(0,X_0)$ cannot be a local but non global minimum for 
$S$ under the natural requirement that each surface $S=$ const. 
corresponds to a unique integral manifold of $\delta Q_{rev}$ [this means that 
isoentropic states are path-connected. Cf. \cite{belhom}]. In fact, 
in homogeneous thermodynamics, there exists an integral manifold 
of $\delta Q_{rev}$ such that $S(B,X)=S(0,X_0)=$ const. if $(0,X_0)$ is not 
a global minimum, and to such an integral manifold $(0,X_0)$ would not 
belong if $(0,X_0)$ is a local minimum. 
Notice also that no point belonging to the boundary $B=0$ can be 
a local maximum, because $S(B,X)$ is a strictly 
increasing monotone function in $B$. 

In any convex neighborhood $W$ of $(0,X_0)$ there exist 
$(B^{+},X^{+})$ and $(B^{-},X^{-})$ such that 
$\sigma (B^{+},X^{+})>0$ and $\sigma (B^{-},X^{-})<0$, because 
$(0,X_0)$ is not a local minimum.  
By continuity, for any convex neighborhood of $(0,X_0)$ there exists 
$(B^{0},X^{0})$ such that $\sigma (B^{0},X^{0})=0$. The point  
$(0,X_0)$ is then a limit point for the set 
$Z(\sigma)\equiv \{(B^{0},X^{0})| \sigma (B^{0},X^{0})=0\}$, which is 
a closed set because $\sigma$ is continuous. In order to show that a solution 
$B(X)$ for $\sigma (B(X),X)=0$ exists and is unique, 
we introduce the following auxiliary function
\beq
\bar{\sigma}(B,X)\equiv \left[
\begin{array}{lcr} 
\sigma (B,X) &   & \hbox{for}\ B\geq0;\cr 
S(0,X)-S(0,X_0)+B &   & \hbox{for}\ B<0.
\end{array} \right.
\eeq 
This function $\bar{\sigma}(B,X)$ is a continuous function 
which is monotone strictly increasing in $B$ also for $B<0$. 
We have extended then $\sigma$ to negative values of $B$, which 
is shown to be an useful trick. We cannot use the 
standard form of the implicit function theorem because $\pa S/\pa B$ 
diverges at $B=0$. Nevertheless, the proof is 
a variant of the standard proof of the implicit function 
theorem for scalar functions. We have that $\bar{\sigma}(0,X_0)=0$.  
By the monotonicity property one has that there exist $B_1<0<B_2$ 
such that $\bar{\sigma}(B_1,X_0)<0<\bar{\sigma}(B_2,X_0)$. 
By continuity, there exists an open neighborhood $R\ni X_0$ such that 
$\bar{\sigma}(B_1,X)<0<\bar{\sigma}(B_2,X)$ for all $X\in R$. 
Then, from 
the intermediate value theorem it follows that there exists a 
value $\bar{B}\in (B_1,B_2)$ such that 
$\bar{\sigma}(\bar{B},X)=0$ for any fixed $X\in R$. 
Monotonicity 
ensures that $\bar{B}$ is unique for each fixed $X\in R$. 
The function $B(X):R\to \RR$ is then 
defined as the map defined by $B(X)=\bar{B}$, where 
$\bar{B}$ is the solution of $\bar{\sigma} (\bar{B},X)=0$ for each fixed 
$X\in R$. Such a function satisfies $B(X_0)=0$ and is also continuous in 
$X$. This 
follows again from the fact that 
$\bar{\sigma}$ is a continuous function which is 
monotonically strictly increasing in $B$. Being continuous, one has that 
the set 
\beq
Z(\bar{\sigma})\equiv \{(B,X)| \bar{\sigma} (B,X)=0\}
\eeq
is a closed set. Given a sequence $\{ X_n \}\subset R$ such that 
$X_n\to \hat{X}\in R$ for $n\to \infty$, one finds that there exists a unique 
(by monotonicity) $\hat{B}$ such that 
$(\hat{B},\hat{X})\in Z(\bar{\sigma})$. This means that 
$\lim_{n\to \infty}\; B(X_n)=B(\hat{X})=\hat{B}$, i.e., $B(X)$ is 
continuous.\\
It is evident that   
\beq
Z(\bar{\sigma})\supseteq Z(\sigma)
\eeq
and we have to get rid of the spurious solution 
$B=-(S(0,X)-S(0,X_0))<0$ which could occur for $S(0,X)-S(0,X_0)>0$. 
But this solution cannot hold for any $X\in R$, because 
in any neighborhood of $(0,X_0)$ there exist points $(0,X^-)$ 
such that $S(0,X^-)-S(0,X_0)<0$. Thus, being the spurious solution 
not defined in the whole $R$, the above theorem allows to 
conclude that $B(X)\geq 0$ surely exists.\\
Notice that, for inner points $B(X)>0$, this solution is 
actually a leaf of the foliation defined by the integrable at least 
$C^1$ Pfaffian form $\omega$. As a consequence, for inner points $B(X)$ 
is at least $C^2$. One can also calculate the gradient of 
$B(X^1,\ldots,X^{n+1})$:
\beq
\left( \frac{\pa B}{\pa X^1},\ldots,\frac{\pa B}{\pa X^{n+1}} 
\right)=\left(\tilde{\xi}_1,\ldots,\tilde{\xi}_{n+1} \right),
\eeq
where the latter equality is due to the fact that $B(X)$ 
satisfies the Mayer-Lie system. It is then evident that 
\beq
\left( \frac{\pa B}{\pa X^1},\ldots,\frac{\pa B}{\pa X^{n+1}} 
\right)\to 0 \quad \hbox{for}\ B\to 0^+,
\eeq
i.e. $B(X)$ reaching $B=0$ is tangent to $B=0$.\\ 

The above condition about the absence of inner integral manifolds 
arbitrarily approaching $T=0$ 
is also sufficient. Notice also that it does not 
forbid the inner leaves to asymptotically approach $T=0$ as some 
deformation variable, say $X^k$, is allowed to diverge:    
$|X^k|\to \infty$ as $B\to 0^+$.  
The unattainability is clearly ensured because of such a divergence. 
Let us consider the following equation:
\beq
\sum_{i=1}^{n+1}\; \tilde{\xi}_i 
(B,X^1(B),\ldots,X^{n+1} (B))\;  \frac{dX^i}{dB}-1=0,
\eeq
which is another rewriting of the above equation where $B$ plays the 
role of independent variable and where $n$ functions $X^i$, say  
$X^1(B),\ldots,X^{n} (B)$, are arbitrarily assigned. For simplicity, 
let us put $X^{n+1}(B)\equiv X(B)$. 
One could also 
consider $X^1=X^1_0,\ldots,X^{n}=X^n_0=$ const. Then, one obtains
\beq
\frac{dX}{dB}=\frac{1}{\tilde{\xi}_{n+1} 
(B,X(B);X^1_0,\ldots,X^{n}_0)}.
\eeq
It is evident from our discussion above that any solution 
$X(B)$ has to be such that 
$|X(B)|\to \infty$ when $B\to 0^+$ if $T=0$ has to be a leaf. 
Some examples are given below. 

\subsubsection{examples}

Let us consider
\beq
\omega=dU+\frac{2}{3} \frac{U}{V} dV;
\eeq
the domain is chosen to be $0\leq U,\ 0<V$ and the 
Pfaffian form $\omega$ is $C^1$ everywhere. 
One has
\beq
f=\frac{5}{3} U
\eeq
which vanishes for $U=0$. The boundary $U=0$ is an 
integral submanifold of $\omega$. Let us consider the Cauchy problem 
\beqnl
\frac{dU}{dV}&=&- \frac{2}{3} \frac{U}{V}\\
U(V_0)&=&0.
\eeqnl
It is evident that the only solution of this problem is 
$U=0$, which is a leaf of the thermodynamic foliation. 
By integrating, one finds the (concave)
entropy $S=c_0\; U^{3/5} V^{2/5}$ and $T=5/(3 c_0)\; (U/V)^{2/5}$ 
[$c_0$ is an undetermined constant]. (N) is satisfied. Along an 
isoentropic surface $S_0>0$, one finds 
\beq
U=\left( \frac{S_0}{c_0} \right)^{5/3}\; V^{-2/3}
\eeq
and $T=0$, i.e. $U=0$ can be approached only for $V\to \infty$.
\\
Let us consider a Pfaffian form having the same domain $0\leq U,\ 0<V$ 
\beq
\omega=dU+\left(\frac{U}{V}\right)^{2/3} dV;
\eeq
the Pfaffian form $\omega$ is not $C^1$ on the boundary $U=0$. 
One has
\beq
f=U+U^{2/3} V^{1/3}
\eeq
which vanishes for $U=0$. The Cauchy problem 
\beqnl
\frac{dU}{dV}&=&- \left(\frac{U}{V}\right)^{2/3}\\
U(V_0)&=&0
\eeqnl
allows two solutions:
\beq
U=0
\eeq
and
\beq
U=({V_0}^{1/3}-V^{1/3}).
\eeq
The latter solution holds for $0<V\leq V_0$, and it can be easily 
identified with the isoentrope $S=S_0=c_0 V_0$, where 
$S=c_0 (U^{1/3}+V^{1/3})^3$ is the (concave) entropy. 
(N) is violated and the two solutions 
are tangent for $U=0$. 

Let us consider the following example, which is inspired to the 
low-temperature behavior of a Fermi gas. The Pfaffian form one 
takes into account is
\beq
\omega = dU+\frac{2}{3}\; \frac{U}{V}\; dV-\left(
-\frac{1}{3}\; \frac{U}{N}+2 c\; \left({N}{V}\right)^{2/3} \right)\; dN,
\eeq
where $c$ is a positive constant. This Pfaffian form is integrable and the 
integrating factor is 
\beq
f=2 U -2 c\; \frac{N^{5/3}}{V^{2/3}}.
\eeq
Then the zero of the integrating factor occurs for 
\beq
U = c\; \frac{N^{5/3}}{V^{2/3}}, 
\eeq
and, by construction, being $f\geq 0$, one imposes $U \geq b(V,N)\equiv 
c\; N^{5/3}/V^{2/3}$. The function $b(V,N)$ is extensive and convex. 
Let us define 
\beq
B= U - c\; \frac{N^{5/3}}{V^{2/3}};
\eeq
This coordinate transformation is regular in $B=0$. We have 
\beqnl
\bar{p}(B,V,N) &=& \frac{2}{3}\; \frac{B}{V}\\
\bar{\mu}(B,V,N) &=& -\frac{1}{3}\; \frac{B}{N},
\eeqnl
and 
\beq
\bar{f} = 2 B.
\eeq
Then, one finds  
\beq
S=\alpha\; B^{1/2} V^{1/3} N^{1/6},
\eeq
($\alpha$ is a proportionality constant) 
which can be easily re-expressed in terms of $(U,V,N)$:
\beq
S=\alpha\; \left( U V^{2/3} N^{1/3}- c N^2 \right)^{1/2}.
\eeq 
Notice that, along $S=S_0=$ const. one has 
\beq
B=\left(\frac{S_0}{\alpha} \right)^2\; \frac{1}{V^{2/3}\; N^{1/3}},
\eeq
which can approach $B=0$ only for $V\to \infty$ and/or $N\to \infty$. 
Moreover, 
\beq
T= \frac{2}{\alpha}\; \left( U V^{2/3} N^{1/3}- c N^2 \right)^{1/2}\; 
N^{-1/3}\; V^{-2/3}
\eeq
and $\pa T/\pa U$ diverges as $T\to 0^+$.

\subsection{conditions for the validity of (N)}

\noindent
We can show, by means of purely thermodynamic considerations, that:\\ 
\\
{\sf in Gibbsian variables, if the homogeneous Pfaffian form 
is (at least) $C^1$ also on the boundary $Z(f)$ and if the entropy 
$S$ is concave, then (N) holds}.\\ 
({\sl sufficient but not necessary condition})
\\
\\
Recall that $S\geq 0$ by construction and also on 
statistical mechanical grounds. 
Let us 
assume that the Pfaffian form $\delta Q_{rev}$ is $C^1$ 
everywhere, also on the boundary, and that it satisfies 
the integrability condition in the inner part of the 
manifold. Then, the integrating factor 
$f$ is a $C^1$ function everywhere.\\ 
In variables $U,V,N$ one has
\beq
f=U+p(U,V,N)\; V-\mu(U,V,N)\; N
\eeq
where $p,\mu$ are $C^1$. Consider
\beq
\frac{\pa f}{\pa U}=1+S\; \frac{\pa T}{\pa U}.
\label{dfu}
\eeq
$S$ cannot be non-negative and concave near 
$T=0$ if it diverges as $T\to 0^{+}$ \cite{landsberg}. 
See also sect. \ref{gibbs}. 
Then concavity and positivity 
also near $T=0$ force the entropy to be finite 
in the limit $T\to 0^{+}$. Moreover,  
one has $\pa T/\pa U=1/C_{X^1,\ldots,X^{n+1}}$, where 
$C_{X^1,\ldots,X^{n+1}}$ is the 
standard heat capacity at constant deformation parameters, 
which has to vanish in the limit 
as $T\to 0^+$ because $S$ has to be finite in that limit. 
As a consequence, $\pa T/\pa U=1/C_{X^1,\ldots,X^{n+1}}\to \infty$ 
as $T\to 0^+$. 
Then, by inspection of (\ref{dfu}), it is evident that 
$f$ is $C^1$ also on the boundary only for $S\to 0^+$ as $T\to 0^+$.\\ 
As a consequence of this theorem, we can conclude that 
any violation of (N) is involved with a Pfaffian form that 
is not $C^1$ also on the boundary, as it can be easily verified 
by considering the examples violating (N) in section \ref{gibbs}.

\subsubsection{geometrical aspects of $f=0$}

If $\delta Q_{rev}$ is at least of class $C^1$ everywhere, then 
$d(\delta Q_{rev})$ is continuous and finite on the 
surface $f=0$ (that is, $T=0$). 
From a geometrical point of view, the surface $f=0$ is 
a so-called {\sl separatrix} according to the definition 
of Ref. \cite{cerveau}, in the sense that the $2-$form 
$(\delta Q_{rev})\wedge df$ vanishes with $f$ on this 
surface. One has $(\delta Q_{rev})\wedge df=-f\; d(\delta Q_{rev})$, 
as it can be easily verified by using (\ref{symm}) and standard 
Cartan identities. 
$f=0$ is also a surface where the symmetry 
associated with the vector field $Y$ becomes trivial \cite{bocharov}, 
in the sense that it becomes tangent to the submanifold 
$f=0$. In the case where the complete integrability 
of $\delta Q_{rev}$ is preserved also on the surface 
$f=0$, one has that $f=0$ appears as a special leaf 
of the thermodynamic foliation, it is indeed the only 
leaf which is left invariant by the action of $Y$.\\  
In general, for a sufficiently regular $\omega$ at $f=0$, 
the property for  $f=0$ to be a separatrix in the sense of 
Ref. \cite{cerveau} ensures that 
the solutions of $\delta Q_{rev}=0$ in the inner part of 
the thermodynamic manifold cannot intersect the integral 
submanifold $f=0$. 
Then, 
the integral manifold $f=0$ 
is a leaf of the foliation defined by $\delta Q_{rev}$ 
if it is also a separatrix. 
Moreover, it is a special leaf, being a separatrix.  
\\
We have shown above that $S\to 0^+$ as $f\to 0^+$ for 
$\omega\in C^1$ everywhere. 
We can note that, if 
the condition for $\delta Q_{rev}$ to be $C^1$ 
also on $f=0$ is relaxed, then the condition 
\beq
|\lim_{f\to 0^+}\; f\; d(\delta Q_{rev})|=0
\eeq
is {\sl not} sufficient for ensuring (N). A simple 
counterexample is given by
\beq
\omega=dU+\left( \frac{U}{V} \right)^{\frac{2}{3}}\; dV
\eeq
where $0\leq U,\ 0<V_1<V$ and where 
$f\; d\omega$ vanishes as $f\to 0^+$ but (N) is violated.\\ 
The requirement that $\omega$ is $C^1$ everywhere 
is, at the same time, too restrictive. In fact, 
it is evident that the following homogeneous Pfaffian form,  
defined for ${\cal D}\equiv \{0\leq U<V/e^2,\quad V>0\}$)
\beq
\omega=dU+p(U,V)\; dV,
\eeq
where 
\[
p(U,V)=\left\{
\begin{array}{lll}
& \frac{U}{V}\; (-1-\log(\frac{U}{V})) &\hbox{if}\ 0<U<V/e^2\cr
&&\cr
& 0 &\hbox{if}\ U=0
\end{array}
\right.
\]
satisfies (N) but is not of class $C^1$ at $T=0$ (the corresponding 
entropy is $S=-V/(\log(U/V))$, which is concave on ${\cal D}$ and can 
be continuously defined to be zero when $U=0$). Moreover, this 
example shows that even the more general setting\\
\\
{\sf $S$ concave, $\frac{\pa f}{\pa U}$ finite as $T\to 0^+$ 
$\Rightarrow$ $S\to 0^+$ as $T\to 0^+$}\\
\\
(which is trivial, because $\pa\; f/\pa\; U=1+S\; (\pa\; T/\pa\; U)$) 
does not correspond to a necessary condition for (N). 

\subsubsection{condition (HOM)} 

In order to give a necessary and sufficient condition for (N), 
we use the following interesting property of $\delta Q_{rev}$. 
We have
\beq
\frac{\delta Q_{rev}}{f}=\frac{dS}{S}.
\label{exfo}
\eeq
Let us consider $\int_{\gamma}\; \delta Q_{rev}/f$, where 
$\gamma$ is a curve having final point at temperature $T$.  
If (N) holds, 
then $\int_{\gamma}\; \delta Q_{rev}/f\to -\infty$ as $T\to 0^+$. 
In fact, if (N) holds, whichever the path $\gamma$ one chooses, 
the integral of $dS/S$ diverges to $-\infty$ as $T\to 0^+$.\\  
If, instead, 
$\int_{\gamma}\; \delta Q_{rev}/f\to -\infty$ as $T\to 0^+$ 
whichever path is chosen for approaching 
$T=0$, then $S\to 0^+$ in the same limit. \\
Then the following theorem holds:\\
\\
{\sf (N) $\Leftrightarrow$ 
$\int_{\gamma}\; \delta Q_{rev}/f\to -\infty$ as $T\to 0^+$ 
whichever path is chosen in approaching $T=0$}.\\
\\
\\
({\sl condition} (HOM) {\sl in the following})\\
\\ 
Notice that, because of the concavity of $S$, one cannot 
have $\int_{\gamma}\; \delta Q_{rev}/f\to +\infty$ as $T\to 0^+$, 
because a non-negative and concave entropy cannot diverge 
(cf. sect. \ref{gibbs}). Thus, once ensured the concavity 
property for $S$ (cf. \cite{belhom}), one has only to check if the 
above integral diverges along any rectifiable curve approaching the 
surface $T=0$. 

Notice that, in this form, the above theorem allows to neglect the 
problem of the actual presence of the boundary $T=0$ in the 
physical manifold. This formulation is also coherent with 
the fact that (N) is formulated as a limit for $T\to 0^{+}$.
\\ 
If (N) is 
violated and the limit $\lim_{T\to 0^+} S$ exists, then 
(\ref{exfo}) is integrable along any path approaching 
$T=0$ with positive entropy (it is 
not integrable along any path approaching $T=0$ with 
vanishing entropy). For example, let us consider the 
following toy-model Pfaffian form
\beq
\omega=dU+
\left(\frac{U}{V}+\alpha_0\; \frac{U^{2/3}}{(V\; N)^{1/3}} \right)\; dV
+\left(\frac{U}{N}+\beta_0\; \frac{U^{2/3}}{(V\; N)^{1/3}} \right)\; dN
\eeq
where $\alpha_0>0,\; \beta_0<0$ are constants and the domain 
is restricted by $V/N\geq -\beta_0/\alpha_0$. Then one has 
\beq
f=3\; U+ \alpha_0\; \frac{(U\; V)^{2/3}}{N^{1/3}}+
\beta_0\; \frac{(U\; N)^{2/3}}{V^{1/3}}.
\eeq
The integrating factor $f$ vanishes as $U\to 0$. $\omega/f$ 
is integrable along any path such that $V/N> -\beta_0/\alpha_0$, 
in fact, if $g(V,N)$ stays for a positive function, one has 
$f\sim U^{2/3}\; g(V,N)$ as $U\to 0^+$. If $V/N= -\beta_0/\alpha_0$, 
then $f\sim 3\; U$ as $U\to 0^+$ and $\omega/f$ is no more integrable 
near $U=0$. Notice that the entropy which corresponds to this 
Pfaffian form is $S=3\; (U\; V\; N)^{1/3}+\alpha_0\; V+
\beta_0\; N\geq 0$. (N) is violated and $S$ vanishes on the 
submanifold $U=0, V/N= -\beta_0/\alpha_0$.
\\ 
Notice that, if 
$S$ is not concave but simply positive, 
then condition (HOM) is still equivalent to (N). 
\\
Notice that (HOM) 
is not affected by the connectedness properties of $f=0$.

\subsection{a sufficient condition for a continuous $S$ 
at $T=0$}. 

The entropy $S$ is required to be continuous at $T=0$. 
This can be obtained 
as follows. It is evident that the continuity of $\has$ implies the 
continuity of $S$. Then, one can impose conditions which allow 
$\has$ to be continuous at $T=0$. Nevertheless, the continuity of 
$\has$ excludes, by direct inspection, the possibility to 
obtain $S=0$ at $T=0$. Thus, one has to find further conditions on 
$\has$ in order to allow the possibility to get a vanishing 
entropy at $T=0$. 
Let us consider 
\beq
\has (\hab,X^1,\ldots,X^{n+1})
=\int_{\hab_0}^{\hab}\; d\hab\; 
\frac{1}{f(\hab,X^1,\ldots,X^{n+1})}+\has (\hab_0,X^1,\ldots,X^{n+1}).
\eeq
We wish to know if the limit as $\hab\to 0^+$ of $\has$ exists. 
A sufficient condition is the following: there exists a positive 
function $\phi(\hab)$ such that 
\beq
\frac{1}{f(\hab,X^1,\ldots,X^{n+1})}< \phi(\hab)\quad \forall\; 
\hab\in (0,\hab_0]\quad \hbox{and}\quad \forall X^1,\ldots,X^{n+1}\in C,
\label{fphi}
\eeq
where $C\subset \RR^{n+1}\cap {\cal D}$ is any open bounded set 
contained in ${\cal D}$ and 
\beq
\lim_{\hab\to 0^+}\; \int_{\hab_0}^{\hab}\; d\hab\; \phi(\hab)<\infty.
\eeq
Then $\int_{\hab_0}^{\hab}\; d\hab\; 1/f$ is uniformly convergent 
and, being $\has$ continuous for $\hab>0$, one finds that 
$\has$ can be extended continuously also at $\hab=0$. This 
condition ensures that (N) is violated. The above condition does not leave 
room for $S=0$ for some (but not all) values of 
$X^1,\ldots,X^{n+1}\in {\cal D}$. 
Actually, continuity on a open bounded 
subset $R\subset \RR^{n+1}\cap {\cal D}$ 
of the allowed values for the variables $X^1,\ldots,X^{n+1}$,  
can be also obtained by assuming that (\ref{fphi}) 
holds on $R$ and not for any open bounded set $C$ 
contained in ${\cal D}$. In this case, 
$\hat{S}$ is continuous at $T=0$ for $X^1,\ldots,X^{n+1}\in R$.\\ 
In order to obtain a condition ensuring (N) 
a sufficient condition is the following: there exists a positive 
function $\phi(\hab)$ such that 
\beq
\frac{1}{f(\hab,X^1,\ldots,X^{n+1})}> \phi(\hab)\quad \forall\; 
\hab\in (0,\hab_0]\quad \hbox{and}\quad \forall X^1,\ldots,X^{n+1}\in C,
\label{fphinf}
\eeq
where again $C\subset \RR^{n+1}\cap {\cal D}$ is any open bounded set, 
and  
\beq
\lim_{\hab\to 0^+}\; \int_{\hab_0}^{\hab}\; d\hab\; \phi(\hab)=-\infty.
\eeq
Then, because the above divergence of the integral is uniform 
in $X^1,\ldots,X^{n+1}$, one finds that $\has=\log (S)\to -\infty$ 
as $T\to 0^+$, i.e. $S\to 0^+$ as $T\to 0^+$. Even in this case, 
one can allow the function $\phi(\hab)$ to be different for different 
subsets whose union covers all the values of the deformation parameters.

\subsubsection{(N): integral criterion}
\label{intecr}

We can consider
\beq
\int_{\Gamma}\; \frac{\omega}{\hat{f}}=\hat{S}(\hab,X^1,\ldots,X^{n+1})
-\hat{S}(\hab_0,X^1,\ldots,X^{n+1});
\eeq
due to the singularity in the integrand in $f=0$, when one considers a 
path approaching $T=0$ the integral has to 
be intended as improper integral. Nevertheless, according to a common use, 
we bypass this specification in the following. We can write 
\beq
\hat{S}(\hab,X^1,\ldots,X^{n+1})
-\hat{S}(\hab_0,X^1,\ldots,X^{n+1})= 
\int_0^1\; d\hab\; 
\frac{a(\hab,X^1,\ldots,X^{n+1})}{\hat{f}(\hab,
X^1,\ldots,X^{n+1})}.
\eeq
We can see that, if $\omega\in C^1({\cal D}\cup \pa {\cal D})$, then 
the above integral diverges 
in such a way that $S\to 0^+$ for $\hab\to 0^+$. In fact, one has 
\beq 
\hat{\xi_i}(\hab,X^1,\ldots,X^{n+1})=
\xi_k(\hab,X^1,\ldots,X^{n+1})-\xi_k(0,X^1,\ldots,X^{n+1})=
\frac{\pa \xi_k}{\pa \hab}(0,X^1,\ldots,X^{n+1})\; \hab+O(\hab^2)
\eeq
and
\beq
\hat{f}=k(0,X^1,\ldots,X^{n+1})\; \hab+O(\hab^2), 
\eeq
where surely $k(0,X^1,\ldots,X^{n+1})>0$ because $\hat{f}\geq 0$. 
As a consequence, near $\hab=0$ the integrand behaves as follows:
\beq
\frac{\omega}{\hat{f}}\sim 
\frac{1}{k\; \hab}\; a\; d\hab.
\eeq
Then the integral diverges as $\log(\hab)$ for $\hab\to 0^+$. 
The function $\phi(\hab)$ is 
\beq
\phi(\hab)=\sup_{(X^1,\ldots,X^{n+1})\in C}\left(\frac{a}{k} 
(0,X^1,\ldots,X^{n+1}) \right)\; 
\frac{1}{\hab},
\eeq
where $C$ is an open bounded set contained in ${\cal D}$.

\subsection{inaccessibility (C) and the failure of (N)}

If (N) holds, the surface $T=0$ is adiabatically inaccessible 
along any adiabatic reversible transformation starting at $T>0$, 
and it is a leaf of a foliation. There is no isoentropic surface 
reaching $T=0$ and the property (U) of unattainability is 
automatically ensured, as well as the principle of adiabatic 
inaccessibility (C).\\

The violation of (N) is instead very problematic from the point of 
view of (C) and of the foliation of the thermodynamic manifold. 
If (N) is violated, $T=0$ is not a leaf and it is possible to reach 
$T=0$ along inner (would-be) leaves $S=$ const. 
Actually, one does not find a  
foliation of the whole thermodynamic domain; if 
$T=0$ is included in the thermodynamic manifold, one finds an  
``almost-foliation'', i.e. a foliation except for a zero-measure 
manifold, in the sense that to the proper inner foliation 
generated at $T>0$ is joined a integral manifold $T=0$ (the adiabatic 
boundary of the thermodynamic domain) 
which breaks the adiabatic inaccessibility, even if only 
along special paths passing through $T=0$. 
In the spirit of the thermodynamic 
formalism, we agree with Einstein's statement that 
the existence of such adiabatic paths is ``very 
hurtful to one's physical sensibilities'' \cite{einstein}. 
It is also evident that the Carnot-Nernst cycle discussed in 
sect. \ref{trasfzero} is allowed, unless 
some discontinuity occurs or the thermodynamic formalism fails 
according to Planck's objection, and that the objections against 
its actual performability can hold only in restricted operative conditions 
(from a mathematical point of view, a path contained in the surface 
$T=0$ is different from a isoentropic path at $T>0$ reaching the 
absolute zero of the temperature). Moreover, the approach to the 
problem by means of $\delta Q_{rev}$ reveals in a straightforward way aspects  
which other approaches cannot easily point out.\\

A further remark is to some extent suggested by black hole thermodynamics, 
where (N) is violated but states at $T=0$ have $S=0$. Cf. \cite{belgbh} 
for a study in terms of Pfaffian forms.  
In order to avoid problems occurring with the surface $T=0$ if 
(N) is violated, one could introduce 
a further hypothesis. 
One could  impose that the entropy is discontinuous 
at $T=0$, and that 
\beq
\Sigma_{0}<\inf_{V,X^1,\ldots,X^n}\; \Sigma(V,X^1,\ldots,X^n).
\label{sinon}
\eeq 
One could then impose that $\Sigma_{0}=0$ for all the systems, 
which would allow to recover an universal behavior. Even the 
adiabatic inaccessibility would be restored, because the second 
law would inhibit to reach $T=0$ adiabatically. 
This behavior characterizes black hole thermodynamics. 
Concavity would be preserved, as well as superadditivity. 
However this choice is arbitrary and even unsatisfactory, 
because a well-behaved foliation of the thermodynamic manifold is 
obtained by hand by means of the discontinuous entropy $S$ just 
constructed. In fact, 
the foliation of the thermodynamic 
manifold, if (N) is violated, is obtained as the union of 
the usual foliation at $T>0$ and a 
special leaf at $T=0$. This foliation is generated by a 
Pfaffian form only in the inner part of the manifold.

\section{Notes on Gibbsian approach}
\label{gibbs}

We recall that in Gibbsian approach \cite{gibbsbook}, 
the existence of the entropy is a postulate, because  
the entropy appears in an axiomatic 
framework. See also Refs. \cite{callen,tisza,beattie}.
In a certain sense, very loosely speaking, Gibbs starts where 
Carath\'eodory leaves \cite{lavenda}. 
This can be considered the reason why in Gibbsian approach 
the problems which can be associated with the surface $T=0$ 
as in the previous section appear to be less evident.\\   
Let us assume the Gibbsian approach to thermodynamics, and 
write the so-called fundamental equation in the entropy representation:
\beq
S=S(U,X^1,\cdots,X^{n+1})
\label{fund}
\eeq
where $X^1,\cdots,X^{n+1}$ are extensive deformation 
variables and $U$ is the internal energy. 
$S$ is required to be a first order positively homogeneous 
function and, moreover, a concave function (for mathematical details 
about convexity we refer to Refs. \cite{rockafellar,roberts}). 
The former property 
ensures the extensivity of the entropy, the latter ensures 
the thermodynamic stability of the system against thermodynamic 
fluctuations. The second law $\Delta S \geq 0$ for an 
insulated system is also ensured.

\subsection{extension of $S$ to $T=0$}

Let us define 
\beq
I(U,X^1,\cdots,X^{n+1})=-S(U,X^1,\cdots,X^{n+1}).
\eeq
In what follows, $x$ stays for a state in the thermodynamic 
manifold: $x\equiv (U,X^1,\cdots,X^{n+1})$. The function 
$I(x)$ is, by definition, a convex function and positively 
homogeneous function. As a consequence, its epigraph is a 
convex cone. 
This convex function $I$ is defined on ${\cal C}$ [we change symbol 
for the domain, what follows holds for a generic convex function 
in a generic convex domain]. 
There is a preliminary problem. One has to define $I(x)$ on the 
boundary $\pa {\cal C}$ and obtain again a convex function. 
This is made as follows \cite{roberts}: $I$ can be extended to the set 
\beqnl
&&{\cal F}= {\cal C}\cup \pa {\cal C}_f,\cr
&&\hbox{where}\cr
&&\pa {\cal C}_f\equiv 
\{y\in \pa {\cal C}\ |\ 
\liminf_{x\to y} I(x)<\infty \}\cr 
&&\hbox{and}\cr
&&I(y)\equiv \liminf_{x\to y} I(x)\quad \forall\; y \in \pa {\cal C}_f. 
\eeqnl
The above extension is convex on a convex set. 
In general, one cannot substitute $\liminf_{x\to y} I(x)$ with 
$\lim_{x\to y} I(x)$ because the latter may not exist \cite{roberts}. 
Moreover, the behavior of the convex 
function $I=-S$ at the boundary has to be such that 
\beq
\liminf_{x\to x_0}\; I(x) > -\infty
\label{bbound}
\eeq
for any $x_0$ belonging to the boundary of the convex domain 
[cf. problem F p. 95 of Ref. \cite{roberts}].\\

In the case of $S$, then the  non-existence of the above limit 
can be considered unphysical. In fact, it can also mean that the 
entropy could approach a different value for the same state 
along different paths starting at the same initial point. In the 
latter case, its nature of state function would be jeopardized, it  
requires 
at least the existence of the limit, that is, the independence of 
the limit from the path chosen. On this topic, see  
in particular \cite{landsberg,landrmp}. In any case, 
the definition offered by the theory of convex functions
\beq
S(y)\equiv \limsup_{x\to y} S(x)\quad \forall\; y \in \pa {\cal C}_f
\eeq
is a rigorous formal prescription, but it is 
not clear to the present author if it could be relevant to the 
physics at hand, if the limit does not exists.\\
Then, we assume that $S$ admits a limit for each point of the 
boundary $T=0$, thus
\beq
S(y)\equiv \lim_{x\to y} S(x)\quad \forall\; y \in \{T=0\}.
\label{lims}
\eeq 
Under this hypothesis, we can extend uniquely $S$ at $T=0$. 
The surface $T=0$ represents (a part of) the boundary $\pa {\cal D}$
for the domain, then it belongs to the 
closure of the convex open set ${\cal D}$. 
A convex set is dense in its closure. As a consequence, a 
continuous function $G$ defined in 
${\cal D}$ can be uniquely extended by continuity on the 
boundary $\pa {\cal D}$ if (and only if), for each point 
$x_b\in \pa {\cal D}$ the limit 
\beq
\lim_{x\in {\cal D}\to x_b}\ G
\eeq
exists.\\
It is still to be stressed that, for a non-negative concave $S$, 
one has to find $\lim_{T\to 0^{+}}\; S<\infty$ as a consequence of 
(\ref{bbound}).

\subsection{attainment of the lower bound of $S$}

Gibbsian approach allows us to conclude 
immediately that, if the upper bound $I_0$ of $I$ is attained, then 
it has to be attained on the boundary of the thermodynamic 
manifold (if a convex function $I$ should get 
a maximum value $I_0$ in a inner point of its convex domain, 
it would be actually a constant function in its domain) \cite{roberts}. 
Then, if the lower bound $S_0$ of $S$ is attained, it is 
attained on the boundary of the domain of $S$. 
Moreover, under very simple hypotheses on the domain, 
the upper bound of $I$ is actually attained \cite{rockafellar,roberts}.   
In particular, it can be attained 
at an extreme point of the boundary. We recall that an extreme point 
of a convex set is a point belonging to the boundary of the set such that 
it is not an inner point of any line segment contained in the set. 
For example, if the set is a closed rectangle, the extreme points are 
the four vertices; if the set is a circle, all the points of the boundary 
(circumference) are extreme points. But notice also that, being the 
thermodynamic domain a convex cone, there is no extreme point apart 
from the origin $0$ of the cone (which cannot be considered a 
physically meaningful state \cite{belhom}).

\noindent
Note that, given the surface $T=0$:
\beqnl
&&T(U,X^1,\ldots,X^{n+1})=0\cr
&&\Leftrightarrow \cr
&&U=U_0(X^1,\ldots,X^{n+1}) 
\eeqnl
then, for each point on this surface, as a consequence of the 
homogeneity of the entropy, it holds
\beq
S(\lambda\; U_0(X^1,\ldots,X^{n+1}),\lambda\; X^1,\ldots,\lambda\; X^{n+1})=
\lambda\; S(U_0(X^1,\ldots,X^{n+1}),X^1,\ldots,X^{n+1}).
\eeq
At the same time, the intensivity (i.e., homogeneity of degree zero) 
of $T$ implies
\beq
T(\lambda\; U_0(X^1,\ldots,X^{n+1}),\lambda\; X^1,\ldots,\lambda\; X^{n+1})=
T(U_0(X^1,\ldots,X^{n+1}),X^1,\ldots,X^{n+1})=0.
\eeq
Then, if $U_0(X^1,\ldots,X^{n+1}),X^1,\ldots,X^{n+1} \equiv X^a_0, 
a=0,\ldots,n$ are the points belonging to the surface $T=0$, 
the cone 
\beq
K_0\equiv \{X^a_0\; |\; \lambda\; X^a_0\in K_0,\; \lambda>0\}
\eeq 
is contained in the surface $T=0$ because of the intensivity of $T$. 
As a consequence, in case of violation of (N), 
one could find a system at $T=0$ having 
an arbitrarily high entropy. Only if $S=0$ at $T=0$ this cannot 
happen, because $S=0$ is a fixed point under scaling of 
the entropy.

\subsection{values of $S$ at $T=0$ and the hypothesis of multi-branching}
\label{multibranch}

We have assumed a continuous $S$ at $T=0$. 
From the point of view of the 
Landsberg's discussion about a multi-branching near $T=0$, 
we have then simply to discuss the following topological 
problem. Is the set $Z(T)$ a connected set? In the case 
it is connected, then we can surely conclude that no 
multi-branching can occur near $T=0$. In fact, the range 
of a continuous function on a connected set is a connected set, 
that is, the range of $S$ at $T=0$ is a connected set contained 
in $\RR$. It has to be an interval  (violation of (N)) 
or a single point (validity of (N)). It is interesting to 
underline that, even if the set $Z(T)$ is not connected, (HOM) 
ensures the validity of (N) and viceversa (there is no possibility 
to find two branches like the ones in Fig. (1a), because 
both have to start at $S=0,T=0$). 
A possibility for getting a multi-branching is 
to violate the concavity at least near $T=0$. For an interesting 
example see Ref. \cite{belgbh}. [Another possibility to get a multi-branching 
could be to consider a system allowing for states at $T<0$, but in this case  
two distinct branches would be found on two different sides of $T=0$].

\subsection{violation of (N) and Landsberg's hypothesis}

\noindent
Consider the following toy-model: 
\beq
S=\gamma_0\; V^{1-\alpha}\; U^{\alpha}+\delta_0\; V
\eeq
where $\gamma_0>0,\delta_0>0$ and $0<\alpha<1$. Then 
\beq
T=\frac{1}{\gamma_0\; \alpha }\; \left(\frac{U}{V}\right)^{1-\alpha} 
\eeq
which vanishes as $U\to 0^+$: $T=0\Leftrightarrow U=0$.  
For the domain let us consider 
${\cal F}=\{U\geq 0\} \cup \{V\geq V_0\}$. $I$ is maximum, that 
is, $S$ is minimum, at the extreme point $(0,V_0)$, as it is 
evident. [Notice that in this example the domain is not a 
convex cone because we introduce a lower bound $V_0$ for $V$, as it 
is physically reasonable in order to justify thermodynamics 
on a statistical mechanical ground. If one considers 
${\cal F}=\{U\geq 0\} \cup \{V>0 \}$, then $\inf (S)=0$, which is  
approached at the only extreme point $(0,0)$ of the cone].  
If $\delta>0$, then (N) is violated and $S$ can assume 
an interval of values at $T=0$. 
If $\delta_0=0$, then (N) is satisfied and $I$ is maximal 
on the line $U=0$. A special case is represented by the photon gas, 
where $\alpha=3/4$.\\ 
This toy-model 
corresponds to the following behavior of $S$ as a function 
of $T$ and $V$:
\beq
S(T,V)=\left( \epsilon_0\; T^{\frac{\alpha}{1-\alpha}}\; 
+\; \delta_0 \right)\; V
\eeq
which, for $\delta_0\not =0$, violates (N). 
It is useful to 
pass to the energy representation
\beq
U=\left( \frac{S-\delta_0\; V}{\gamma_0} \right)^{\frac{1}{\alpha}}\; 
V^{\frac{\alpha-1}{\alpha}}.
\eeq
The domain is 
${\cal G}=\{V_0\leq V\leq S/\delta_0\}$. The 
$T=0$ surface corresponds to $V=S/\delta_0$. 
We have 
\beq
T=\frac{1}{\alpha}\; \frac{1}{ \gamma_0^{\frac{1}{\alpha}} }\; 
V^{ \frac{\alpha-1}{\alpha} }\; 
\left( S-\delta_0\; V \right)^{ \frac{1}{\alpha}-1 }; 
\eeq
the pressure is 
\beq
p=\frac{1}{\alpha}\; \left(\frac{1}{\gamma_0}\right)^{\frac{1}{\alpha}}\; 
\left( S-\delta_0\; V \right)^{\frac{1}{\alpha}-1}\; 
V^{-\frac{1}{\alpha}}\; (S\; (1-\alpha)\; +\; V\; \alpha\; \delta_0).
\eeq
The isoentropic $S=S_0$ has equation 
\beq
U(V)=\left( \frac{S_0-\delta_0\; V}{\gamma_0} \right)^{\frac{1}{\alpha}}\; 
V^{\frac{\alpha-1}{\alpha}}
\eeq
and reaches $T=0$ when $V=S_0/\delta_0$ during an adiabatic expansion. 
It is easy to see that it is tangent to the $T=0$ surface. 
The adiabatic expansion has to stop there, because of the structure 
of the domain. One can wonder if any physical reason for such a 
stopping exists. It is useful to come back to Landsberg's suggestion 
about a possible vanishing of the adiabatic compressibility: 
\beq
K_S=-\frac{1}{V}\; \left(\frac{\pa V}{\pa p}\right)_S.
\eeq
In our case, we get
\beq
K_S=\alpha^2\; \gamma_0^{\frac{1}{\alpha}}\; \frac{1}{1-\alpha}\; 
\frac{1}{S^2}\; V^{\frac{1}{\alpha}}\; 
(S-\delta_0\; V)^{\frac{2\alpha-1}{\alpha}}
\eeq 
and three cases occur: when $1/2<\alpha<1$ then $K_S\to 0$ as 
$T\to 0^+$, in such a way that the elastic constants of the system 
diverge in that limit, forbidding any further expansion (Landsberg's 
behavior); when $0<\alpha<1/2$ then $K_S\to \infty$ as 
$T\to 0^+$, the elastic constants vanish and the behavior of the system 
is pathologic (the system appears to be ``totally deformable'' 
in that limit); when $\alpha=1/2$ then 
$K_S\to \gamma_0^2\; V^2/(2\; S^2)$ which is in any case finite 
($S$ is surely positive and not zero) and a physical hindrance 
against reaching $T=0$ is not apparent. 
For $\alpha=3/4$ the behavior for $S$ as 
$T\to 0^+$ is like the one of Debye model, except for the shift 
$\delta_0 V$ which allows the violation of (N); the adiabatic 
expansion stops at $T=0$ because of a vanishing $K_S$. For 
$\alpha=1/2$ one obtains a behavior similar to the one of an 
electron gas near $T=0$, except for the shift $\delta_0 V$; 
no vanishing or divergence of $K_S$ is allowed at $T=0$.\\
We wish to underline that, if one consider the entropic  
fundamental equation of an electron gas at low temperature 
and shifts it by $\delta_0\; V$, 
then the situation is still different, 
because of the zero-point mode contribution. In fact, such a 
contribution allows to get a positive and 
finite $p$ as $T\to 0^+$ and, moreover, a positive and finite 
$K_{S,N}$ as $T\to 0^+$ (the domain is 
$V> 0,N> 0,U\geq U_0 (V,N)$, where $N$ is the particle 
number). 
As a consequence, Landsberg's mechanism does not seem to be 
available. Instead, even if the zero-point mode 
contribution is taken into account in the case of a Debye crystal, 
the pressure is still positive as $T\to 0^+$ but 
$K_{S,N}$ vanishes.

\subsection{further properties}

Another point that can be underlined is the following. 
Let us assume to extend the convex function $I=-S$ to 
all of $\RR^n$, by defining $I=+\infty$ outside its domain  
$\hbox{dom}\; I$. Then, replace this function with its closure. 
This is the same procedure which is prescribed in Ref. \cite{terhorst} 
for the internal energy $U$. 
Then $I$ is a closed proper convex function which is 
{\sl essentially smooth}, that is, $|\nabla\; I|\to \infty$ for 
any subsequence converging to a boundary point. In fact, 
in the gradient of $I$ appears the factor $1/T$ which diverges 
as the boundary $T=0$ is approached. This allows to obtain 
in thermodynamics a convex function of Legendre type, which is 
relevant for the discussion of Legendre transformations 
in thermodynamics. The difference of the entropy 
representation with respect to the energy representation 
is evident from this point of view. In fact, the opposite conclusion 
appears in Ref. \cite{terhorst} for the energy representation: 
no Legendre type function exists in thermodynamics. But we 
think, from the discussion of the previous section, that 
the entropy representation is more fundamental with respect 
to the energy representation at least for the discussion 
of the boundary $T=0$.\\ 

Summarizing our analysis in Gibbs framework:\\
\\
g1) if $S\geq 0$ , then $S=0$ can be 
attained at a point of the boundary of the domain;\\ 
g2) it could be that $S>0$ at other points on the surface 
$T=0$. Then, homogeneity implies that, by scaling, a system 
with an arbitrarily high zero temperature entropy could be 
obtained;\\
g3) models exist where the violation of (N) does not imply the 
attainability of $T=0$ and the violation of $\Delta S=0$ for 
adiabatic reversible transformations (states at $T=0$ are not 
available and so no such violation can occur at $T=0$). But 
these systems display a non-universal behavior (i.e., a behavior 
which does not appear in other models).

\section{conclusions}
\label{concl}

We have discussed the status of third law of thermodynamics 
and we have given an heuristic argument in favor of the entropic 
version of the third law. Then, we have analyzed the law both 
in Carath\'eodory's approach and in Gibbs' approach 
to thermodynamics. 

In particular, Carath\'eodory's approach shows that for $T>0$
the thermodynamic manifold can be foliated into leaves 
which correspond to isoentropic surfaces. The only 
hypothesis is that the Pfaffian form $\delta Q_{rev}$ 
is integrable and $C^1$ in the inner part ($T>0$) of the 
thermodynamic manifold. At the boundary $T=0$, 
which is assumed to be an integral manifold of the Pfaffian 
form $\delta Q_{rev}$, the aforementioned 
Pfaffian form is allowed to be also only continuous.  
The special integral manifold 
$T=0$ is problematic from a physical point of 
view, because it can also be intersected by the inner 
(would-be) leaves $S=$ const. In the latter case, (N) is 
violated and one obtains an almost-foliation of the 
thermodynamic manifold, where the inaccessibility property 
fails, even if only along special adiabatic paths which 
pass through the surface $T=0$. For an entropy which is continuous  
also at $T=0$, (N) holds if and only if $T=0$ is a leaf. 
This is a remarkable result, the validity of (N) is strongly related to 
the possibility to obtain a foliation for the whole 
thermodynamic manifold, including $T=0$.  
We have shown that, if 
the Pfaffian form is $C^1$ everywhere, then (N) is preserved. 
Physical assumptions and mathematical conditions have been 
discussed. 

In another paper \cite{belg31}, further conditions 
leading to the third law are discussed.  

We add herein some notes about the conditions ensuring (N) in 
quasi-homogeneous thermodynamics introduced in \cite{qotd}. 
Also in the quasi-homogeneous case (N) holds iff $\lim_{T\to 0^+}\; S=0$. 
The analysis of sects. \ref{bou} and \ref{leafzero} hold with obvious 
changes;  moreover, condition (HOM) holds unaltered, and, if the Pfaffian 
form $\omega$ is $C^1$ everywhere, then (N) holds (this can be shown by 
using a criterion analogous to the one appearing in (\ref{intecr}). 

\vskip -0.4truecm

\acknowledgments

The author wishes to thank Lawrence Conlon for his clarifying e-mail 
on Frobenius theorem in presence of manifolds with boundary.

\end{document}